\documentclass[a4paper, amsmath, amssymb, showpacs, superscriptaddress, preprintnumbers, onecolumn, 12pt]{revtex4-1}
\usepackage{amsmath}

\usepackage{graphicx}
\usepackage{epstopdf}
\usepackage{mathptmx, textcomp}
\usepackage[latin1]{inputenc}
\usepackage[T1]{fontenc}
\usepackage{natbib}
\usepackage[export]{adjustbox}
\usepackage[american]{babel}
\usepackage[left=2.0cm,top=2.5cm,right=2.0cm,bottom=2.5cm,nohead]{geometry}
\usepackage{fancyhdr}
\usepackage{color}
\usepackage{upgreek}
\fancypagestyle{plain}{
\fancyfoot[C]{\thepage}}
\usepackage{bm}
\bibpunct{(}{)}{,}{n}{,}{,}
\DeclareMathAlphabet{\mathcal}{OMS}{cmsy}{m}{n}

\begin{document}

\title{State-to-state chemistry at ultra-low temperature}
\ \\

\author{Joschka Wolf}
\affiliation{Institut f\"{u}r Quantenmaterie and Center for Integrated Quantum Science and Technology IQ$^{ST}$, Universit\"{a}t Ulm, 89069 Ulm, Germany.}

\author{Markus Dei{\ss}}
\affiliation{Institut f\"{u}r Quantenmaterie and Center for Integrated Quantum Science and Technology IQ$^{ST}$, Universit\"{a}t Ulm, 89069 Ulm, Germany.}
\author{Artjom Kr\"{u}kow}
\affiliation{Institut f\"{u}r Quantenmaterie and Center for Integrated Quantum Science and Technology IQ$^{ST}$, Universit\"{a}t Ulm, 89069 Ulm, Germany.}
\author{Eberhard Tiemann}
\affiliation{Institut f\"ur Quantenoptik, Leibniz Universit\"at Hannover, 30167 Hannover, Germany.}
\author{Brandon P. Ruzic}
\affiliation{Joint Quantum Institute, University of Maryland and NIST, College Park, MD 20742 USA.}
\author{Yujun Wang}
\affiliation{American Physical Society, 1 Research Rd., Ridge, NY 11961 USA.}
\author{Jos\'{e} P. D'Incao}
\affiliation{JILA, NIST and Department of Physics, University of Colorado, Boulder, CO 80309-0440 USA.}
\author{Paul S. Julienne}
\affiliation{Joint Quantum Institute, University of Maryland and NIST, College Park, MD 20742 USA.}
\author{Johannes Hecker Denschlag$^*$}
\affiliation{Institut f\"{u}r Quantenmaterie and Center for Integrated Quantum Science and Technology IQ$^{ST}$, Universit\"{a}t Ulm, 89069 Ulm, Germany.}

\begin{abstract}
\vspace{0.5cm}
\normalsize{$^\ast$ E-mail:  johannes.denschlag@uni-ulm.de.}
\vspace{1cm}
\textbf{\\
Fully understanding a chemical reaction on the quantum level is a long-standing goal in physics and chemistry. Experimental investigation of such state-to-state chemistry requires both the preparation of the reactants and the detection of the products in a quantum state resolved way, which has been a long term challenge. Using the high level control in the ultracold domain, we prepare a few-body quantum state of reactants and demonstrate state-to-state chemistry with unprecedented resolution.  We present measurements and accompanying theoretical analysis for the recombination of three spin-polarized ultracold Rb atoms forming a weakly bound Rb$_2$ dimer. Detailed insights of the reaction process are obtained that suggest propensity rules for the distribution of reaction products. The scheme can readily be adapted to other species and opens a door to detailed investigations of inelastic or reactive processes in domains never before accessible.}\end{abstract}

\maketitle
 Although the underlying fundamental forces and equations are well known, there is no full understanding of the inelastic or reactive dynamics of a system with more than two reactive constituents. Solving the complete dynamics is only possible when the number of product exit channels is very limited \cite{Suno2009,Wang2011b,Croft2017,Wang2011}.
 In a typical chemical reaction, however, where hundreds of molecular states are involved, a direct solution is currently beyond reach. So far, experimental investigations of the state-to-state processes have been limited because of the challenges to prepare the reactants in a well-defined quantum state and to detect the products in a quantum state resolved manner. Early experiments investigated the exchange collision between molecular and atomic hydrogen, see e.g. \cite{Gerrity1984, Neuhauser1992}. These experiments took place at collision energies of about 1$\:$eV and resolved vibrational and rotational levels of the product molecule, as well as the angular distribution \cite{Kitsopoulos1993}. Another example for state-to-state experiments are half-collisions where a molecule is prepared in a well-defined predissociation state and its fragments are detected after dissociation, see e.g. \cite{Becker1995}. Recently, in the ultracold regime systems with a single reaction channel have been investigated, e.g. \cite{Weber2003,Jochim2003,Rui2017}.

Three-body recombination is one of the most fundamental processes and is highly relevant throughout many different fields, including plasma physics \cite{Lyon2017}, combustion and atmospheric\\ chemistry \cite{Baulch1992,Brown1999}, and primordial star formation \cite{Turk2011, Forrey2013}. Recombination is an exothermic reaction, where three atoms collide to form a diatomic molecule and a free atom, both carrying away the released energy in the form of kinetic energy \cite{Moerdijk1996,Fedichev1996,Soeding1999}.
The reaction can strongly depend on the collision energy, the initial quantum state of the reactants and the details of the interactions between the particles. Ultracold atomic gases allow for extraordinary control of these parameters \cite{Chin2010,Burt1997,Quemener2012},
enabling studies that determine the scaling properties of the total three-body recombination, see e.g. \cite{Esry1999,Bedaque2000,Petrov2003,Petrov2004,Incao2005,Kraemer2006,Braaten2006,Haerter2012,Wang2014,Shotan2014,Perez2014,Kruekow2016}.
In principle, there is a wide range of molecular states that can be formed, ranging from weakly to deeply bound.
The investigation of the final state product distribution is essentially still terra incognita.

 In order to form a molecule with a given bond-length, the three atoms need to approach each other within roughly one bond-length. Since for dilute gases three-body collisions happen more frequently at large distances than at short ones, the formation of weakly bound molecules which have large bond-lengths is favored \cite{Perez2014}. At ultracold temperatures the scattering length sets a typical length scale for the minimal distance between the particles as they pass each other in a collision. Indeed, early investigations of three-body recombination showed that for resonant interactions (large scattering lengths) almost exclusively molecules form in the most-weakly bound state, which has a typical size given by the scattering length \cite{Jochim2003,Weber2003}. For non-resonant interactions (small scattering lengths) also more-deeply bound molecular states are expected to be populated, and it is an open question how sensitive the product distribution is on the details of the interaction potential \cite{Gonz2014,Nesbitt2012}.

Here, we investigate three-body recombination for non-resonant interactions with a gas of ultracold $^{87}$Rb atoms confined in a crossed optical dipole trap (see \cite{Supp}). Each atom is in the electronic ground state with total angular momentum and magnetic quantum numbers $f=1$ and $m_f=-1 $, respectively. Three-body recombination produces Rb$_2$ molecules in the mixed $X^1\Sigma_g^+$ and $a^3\Sigma_u^+$
 electronic state manifolds. We measure the product state distribution down to binding energies of 17$\:\text{GHz}\times h$ identifying hyperfine states with an unprecedented resolution of about 5 $\:\text{MHz}\times h$. Our measurements and calculations indicate that the most-weakly bound molecular level is populated the most ($\approx50\%$), and the population of more-deeply bound levels decreases slowly with binding energy, shedding light on open questions about product distributions. Furthermore, propensity rules can be deduced, e.g. that a newly formed weakly bound molecule inherits the internal spin-quantum numbers of the initially prepared atomic scattering state. Thus, the molecule formation process does not involve spin flips. In contrast to that, we observe that the product molecule can pick up a sizeable rotational angular momentum of up to 6$\hbar$.
\begin{figure*}[h!]
	\begin{center}
		\includegraphics[width = 1\textwidth]{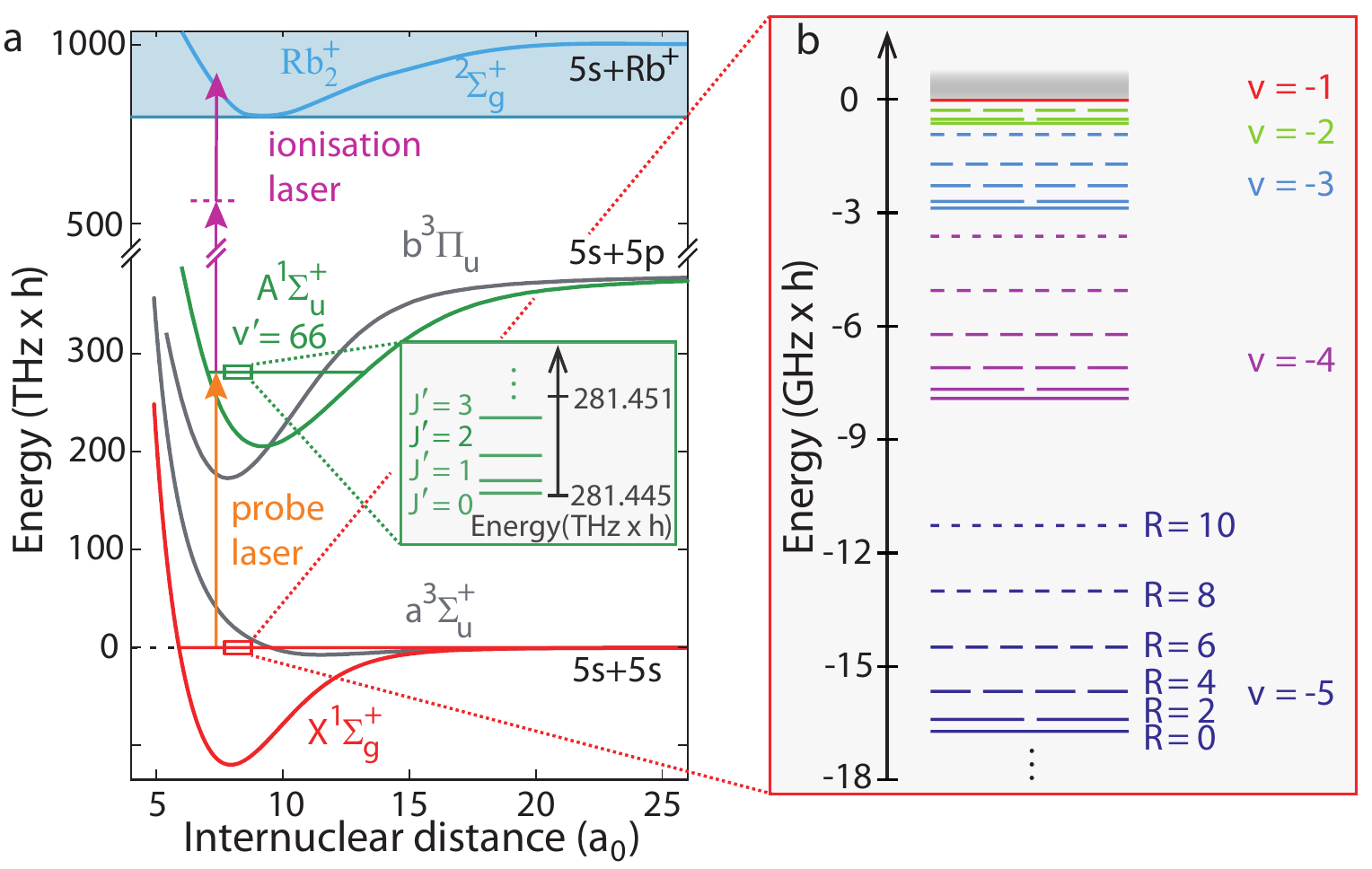}
	\end{center}
	\renewcommand{\figurename}{\textbf{Fig. 1:}}
	\renewcommand{\thefigure}{\hspace*{-2.1 mm}}
	\renewcommand{\labelsep}{none}
	\caption{\textbf{REMPI scheme and overview of relevant molecular states.} \textbf{A},\:A two-color (1,2) REMPI scheme detects weakly bound molecules close to the $5s+5s$ dissociation threshold. The probe laser drives a resonant transition towards the $A^1\Sigma_u^+$, $\textrm{v}'=66$ vibrational level which exhibits a simple rotational substructure  (see inset). $J'$ is the total angular momentum quantum number excluding nuclear spins. Afterwards, two photons from the ionization laser ionize the molecule.  $a_0$  is the Bohr radius.
  \textbf{B},\: Calculated energy levels of selected, weakly bound molecular states with the quantum numbers $\textrm{v}$, $R$ for vibration and mechanical rotation, respectively. Only levels with total positive parity and angular momentum $F=2$ that correlate to the $f_a = f_b = 1 $ atomic asymptote are shown. This asymptote marks the zero energy reference level. The vibrational quantum number $\text{v}$ is counted downwards starting at $\textrm{v} = -1$ for the most-weakly bound vibrational state. }
	\label{fig:Fig1}
\end{figure*}

\begin{section}*{Detection scheme for molecular states}	
For detecting the molecular quantum states we ionize the newly formed molecules in a state-selective fashion using a two-color (1,2) REMPI technique (resonance-enhanced multi-photon ionization).
 Subsequently, the ions are captured in a Paul trap. After a given time, during which several ions can accumulate, we measure their number.\\
Figure$\:$1A shows the REMPI scheme.
 By tuning the frequency of the 'probe laser' (see \cite{Supp}) weakly bound molecules can be state-selectively excited towards the vibrational level $\textrm{v}'=66$ of the $A^1\Sigma_u^+$ potential in a resonant way (for selection rules  see \cite{Supp}).
This level exhibits a simple rotational substructure \cite{Drozdova2013, Deiss2015} (see inset) with predicted hyperfine splittings of less than 3$\:\text{MHz}$.
 From there, two photons of an \grq ionization laser' ionize the molecule (see \cite{Supp}).
  In Fig.$\:$1B we plot the term energies of the most relevant weakly bound molecular levels with $F=2$ for the atom pair  $f_a=1, f_b=1 $, where the indices represent atom $a$ and $b$, and $F$ denotes the total angular momentum of the molecule excluding its rotational angular momentum $R$. The term energies are obtained from coupled-channel calculations,  see \cite{Supp}.
\end{section}

\begin{section}*{Assignment of product states}
 As a  preparation for the assignment of product states we first calibrate the probe laser frequency with respect to the term energies of the probed  weakly bound levels. For this, we carry out a photoassociation measurement \cite{Jones2006} which sets a marker for  zero binding energy.
The ion trap is turned off and the probe laser frequency $\nu$ is scanned in small steps of 5 MHz. For every step we expose a freshly prepared atom cloud to the probe laser for a duration of $2\:\text{s}$ and afterwards measure the remaining number of atoms $N$ via absorption imaging.
On resonance the probe laser couples two ultracold Rb atoms colliding in a $s$-wave (i.e. $R=0$) to the bound level $A^1\Sigma_u^+$, $\text{v}' = 66$, $J' = 1$ and atom loss occurs. Here, $J'$ is the total angular momentum quantum number excluding nuclear spins.
As shown in Fig.$\:$2 (orange data), we observe a single photoassociation line at a frequency of $\nu_0=281,445.045\:\textrm{GHz}$.
The linewidth of the photoassociation dip is $15\:\text{MHz}$, close to the natural linewidth of about $10\:\text{MHz}$.
The next photoassociation line is expected 570 GHz away \cite{Deiss2015}.
\begin{figure}[h!]
	\begin{center}
		\includegraphics[width = 1\textwidth]{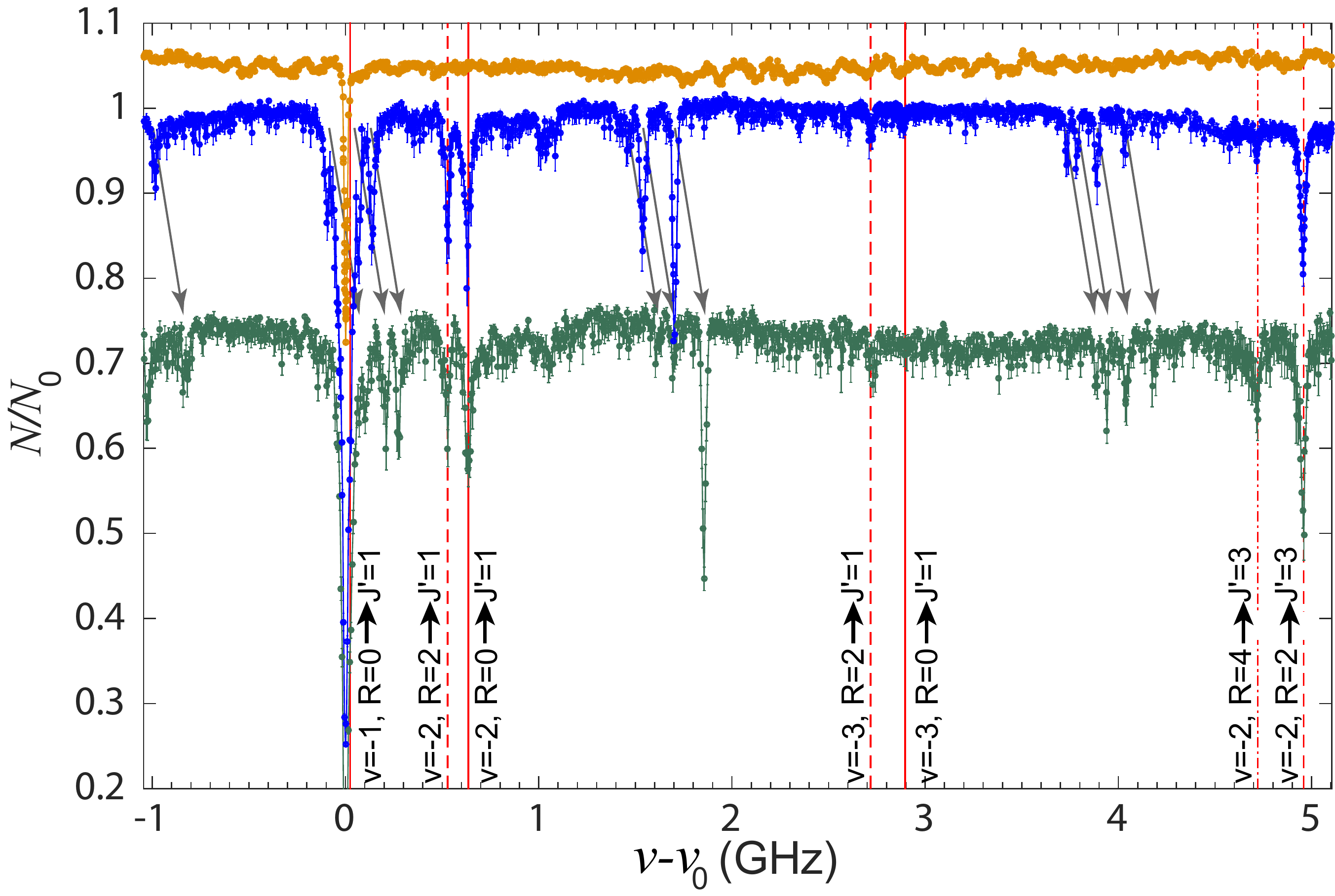}
	\end{center}
		\renewcommand{\figurename}{\textbf{Fig. 2:}}
	\renewcommand{\thefigure}{\hspace*{-2 mm}}
	\renewcommand{\labelsep}{none}
	\caption{\textbf{Photoassociation and REMPI spectra.}
Shown is the remaining atom fraction $N/N_0$ as a function of the probe laser frequency $\nu$, where
 $N_0$ is the number of remaining atoms for a far off-resonant probe laser.
 Orange data correspond to the photoassociation spectrum, with a single line at $\nu = \nu_0 \equiv 281,445.045\:\textrm{GHz}$. For better visibility the record is shifted up by 0.05 units. The blue and green data are REMPI spectra. For the blue data the ionization laser frequency is $281,629.15\:\textrm{GHz}$, while for the green data it is 150$\:$MHz smaller.
Each REMPI data point is the average of 10 repetitions with the error bars being the statistical standard deviation. For better visibility the green spectrum is vertically shifted by $-$0.25 units, which cuts off part of its photoassociation line. The vertical lines mark assigned resonant transitions of the first REMPI step. For grey arrows see text.}
	\label{fig:Fig2}
\end{figure}

We now turn on the ion trap and repeat the experiment with an exposure time of $0.5\:\text{s}$ (blue data). In addition to the strong photoassociation dip at $\nu = \nu_0$ also a number of other loss dips are observed. As our analysis will show, these resonances belong to product molecules from three-body recombination which are state-selectively ionized as the probe laser is scanned. The ionized molecules are immediately trapped in the Paul trap which is arranged such that they are immersed in the cold atom cloud. The ions undergo collisions with the atoms and induce atomic losses that can be substantial during exposure time. A larger number of ions leads to a larger loss of atoms. Resonance dips in Fig.$\:$2 with a depth of about 0.3 are typically induced by an average of about five ions. This measurement of atomic loss constitutes a semi-quantitative ion detection method which we refer to as ion detection scheme I (see \cite{Supp} for details).

 We first consider REMPI transitions towards $A^1\Sigma_u^+$, $\text{v}' = 66$, $J' =  1$ and turn to $J' > 1$ afterwards.
As the probe laser frequency increases, starting from $\nu_0$, it probes more and more-deeply bound molecular levels. In the given frequency range of Fig.$\:$2 we observe signals from the most-weakly bound vibrational states $\text{v} = -1, -2, -3$. The five vertical lines at $\nu-\nu_0< 3\:\text{GHz}$  show the predicted resonance positions for the transitions as labeled next to the lines, probing weakly bound states characterized by $f_a = f_b = 1, F = 2, R= 0$ and $ 2 $ (see also Fig.$\:$1B). These predicted lines match very well with our measured resonances within a few MHz. Some of these  molecules have  non-vanishing rotational angular momentum $R$ (see e.g. $R=4$ in Fig.$\:$2). This is remarkable, since at ultracold temperatures the atoms originally collide in $s$-waves where $R = 0$.

As a consistency check of our assignment, we make use of the fact that for molecules with $R > 0$ each level can give rise to two transition lines, $R \rightarrow J' = R\pm 1$. Indeed, we verify that e.g. the level $\text{v} = -2, R = 2$ not only produces a transition line at $\approx 0.5\:\text{GHz}$ ($J' = 1$) but also one at $\approx 5\:\text{GHz}$ ($J' = 3$) (see Fig. 2). Both transition lines are of similar strength, as expected.

After determining that the observed resonance lines belong to weakly bound molecules in well defined quantum states, one might still question whether the origin of the molecules is three-body recombination. We have performed test measurements on the density dependence of the ion signal for transition lines (see Fig.$\:$S1 in \cite{Supp}). The normalized ion signals show a clear quadratic scaling behavior which points to the three-body nature of the molecule formation.

Besides the weakly bound molecular states with the quantum numbers $f_a = f_b = 1$,  there exist other states near threshold with the quantum numbers $f_a = f_b = 2$ or $f_{a/b} = 1, f_{b/a} = 2$, see \cite{Supp}.
However, we only find clear signals from $f_a = f_b = 1$ molecules in our measurements.  Furthermore, we only observe molecules with an even rotational quantum number, i.e. $R = 0, 2, 4,...$, which corresponds to positive total parity.  Interestingly, the spin quantum numbers $f_a$ and  $f_b$ as well as the total parity of the product molecules are the same as for the two-body atomic scattering state initially prepared in our experiment. The total parity of the scattering state must be even because the colliding atoms are identical bosons and hence their partial wave must be even. Thus our present experiments indicate that the internal spin states of the colliding atoms do not change when a weakly bound molecule is formed in three-body recombination. This is in contrast to our previous measurements where we investigated more-deeply bound Rb$_2$ molecules  and observed a broad range of spin states \cite{Haerter2013}.

In addition to the assigned transition lines  several other unidentified resonance dips are visible in Fig.$\:$2 (blue data). In order to investigate where these lines originate from we measure the spectrum again, but with the ionization laser frequency shifted by $-150\:\text{MHz}$ (green data in Fig.$\:$2). Both scans exhibit the already assigned transition lines at the same probe laser frequencies. The positions of several unidentified resonance dips, however, shift by $+150\:\text{MHz}$. Thus, for these dips, the sum of the ionization and probe laser frequencies remains constant. This indicates a two-photon process, where the combination of a probe and an ionization photon resonantly excites an intermediate molecular level. A third photon finally ionizes the molecule. Therefore, these transition lines correspond to a (2,1) REMPI process. Currently, we have not yet assigned these lines to specific molecular transitions. We note that further unidentified resonance dips can be found in the spectrum, which, however, belong to more deeply bound states with other REMPI paths and need to be discussed elsewhere.

\end{section}

\begin{figure}[t]
	\begin{center}
		\includegraphics[width = 0.55\textwidth]{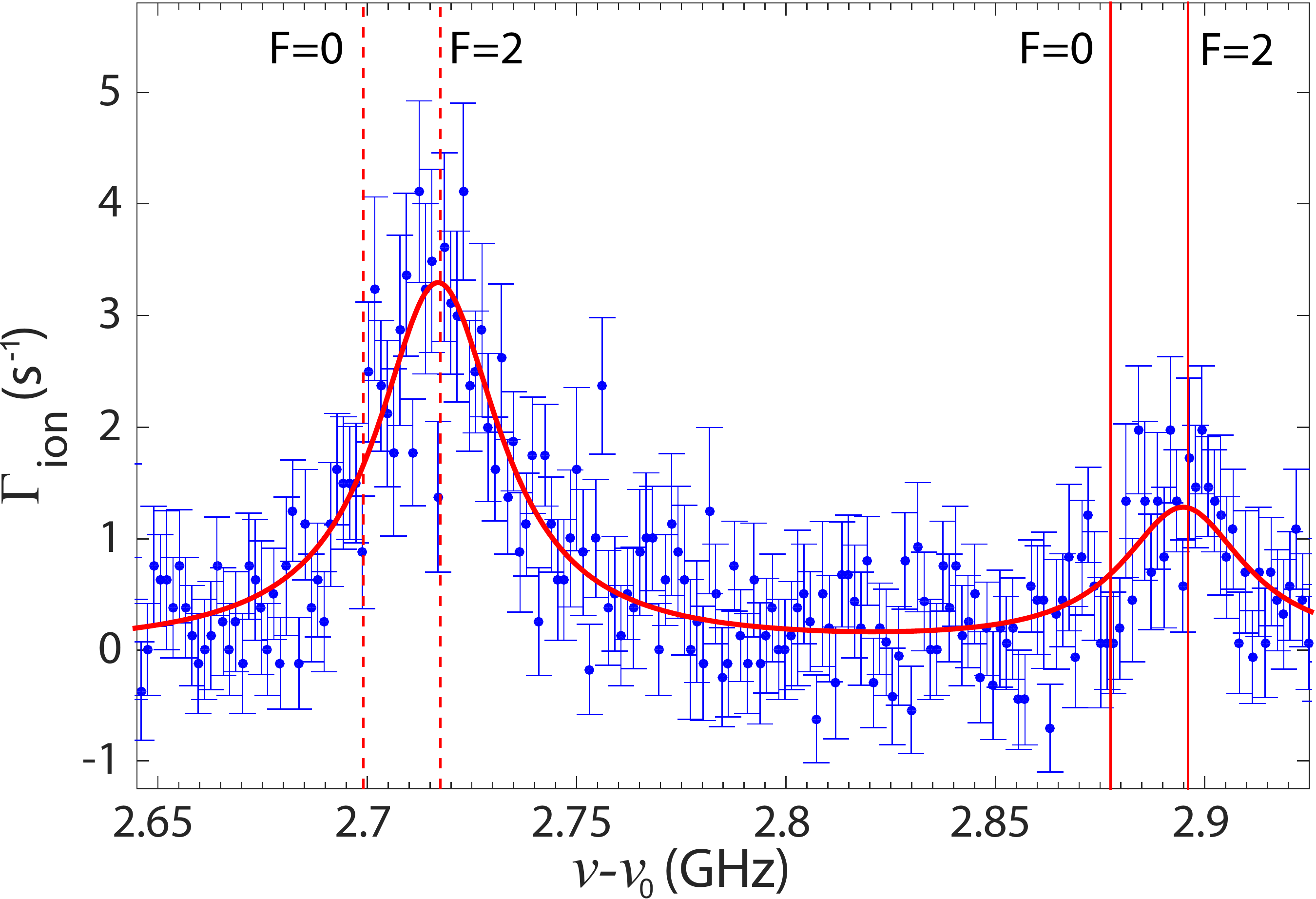}
	\end{center}
		\renewcommand{\figurename}{\textbf{Fig. 3:}}
	\renewcommand{\thefigure}{\hspace*{-2mm}}
	\renewcommand{\labelsep}{none}
	\caption{\textbf{Discrimination of hyperfine levels.}
This REMPI spectrum  shows two transition lines towards $J'=1$ starting from $\textrm{v}=-3,R=0$ (peak on the right) and $\textrm{v}=-3,R=2$ (peak on the left), respectively (see also Fig.$\:$2).  $\Gamma_\mathrm{ion}$ is the ion production rate. The vertical lines show calculated positions of possible product signals with hyperfine quantum numbers $F=0$ and $2$. The data reveal that only $F=2$ states are significantly populated. Each data point is the average of 43 repetitions, and error bars indicate the statistical standard deviation. The red solid line is a Lorentzian fit of the two transition lines. As before, $\nu_0=281,445.045\:\textrm{GHz}$.
}
	\label{fig:Fig4}
\end{figure}
\begin{section}*{Resolution of hyperfine product states}

Our spectroscopic method can resolve molecular levels well beyond the vibrational and rotational splitting, revealing details of hyperfine and exchange interaction.
Figure $\:$4 shows the transitions $\textrm{v}=-3, \, R=0,\, 2 \rightarrow J' = 1$, observed with ion detection scheme II
 which accurately measures the number of trapped ions (see \cite{Supp}). Each of the $R = 0, \, 2$ levels consists of two ($F=0$ and $F=2$) hyperfine sub-levels with a small splitting of about $20\:\textrm{MHz}$. (The substructure of $F$ levels is negligible, see \cite{Supp}.). The small splitting is linked to the fact that the singlet $X^1\Sigma_g^+$ and triplet $a^3\Sigma_u^+$ potentials have slightly different scattering lengths. The calculated transition frequencies are indicated as vertical lines in Fig.$\:$4.
 In the spectrum we only observe hyperfine levels with $F=2$. From the peak-heights and the noise we estimate that residual $F=0$ signals must be at least a factor of four smaller as compared to $F=2$. In fact, also in all other measurements for weakly bound molecules we only find signatures for levels with total internal spin $F=2$. Interestingly, $F$ is the same as for the scattering state.  Thus, this finding supports our previous hypothesis
that the molecular product has the same internal spin state as the two-body atomic scattering state.
\end{section}

\begin{section}*{Most-weakly bound state}

Next, we investigate the most-weakly bound molecular level with positive parity ($\textrm{v}=-1,R=0$) since it is expected to be dominantly populated in ultracold three-body recombination. This level has a binding energy of $24\:\textrm{MHz}\times h$. Unfortunately, the REMPI signal of this molecular level is buried under the strong photoassociation line in Fig.$\:$2. Therefore, we enhance the photoexcitation signal relative to the photoassociation line by reducing the probe laser intensity by a factor of 20 to $\approx0.6\:\textrm{W\,cm}^{-2}$.
This diminishes saturation effects of the photoexcitation peak and results in a narrow linewidth of about $16\:\textrm{MHz}$ (FWHM), close to the natural linewidth of $\approx10\:\textrm{MHz}$ for the $A^1\Sigma_u^+$ state. Figure$\:$3 shows the ion signal  of the photoassociation and photoexcitation lines  measured with ion detection scheme II. The photoexcitation line is located at $\nu = \nu_0 + 24\:$MHz, on the shoulder of the large photoassociation peak. From a fit of two Lorentzians to the lineshape we  extract the signal strength of the photoexcitation peak. Our analysis in the next section shows that about $50$\% of all  molecules that are formed in three-body recombination are produced in the most-weakly bound level at $24\:\text{MHz}$.
\begin{figure}[t!]
	\begin{center}
		\includegraphics[width = 0.55\textwidth]{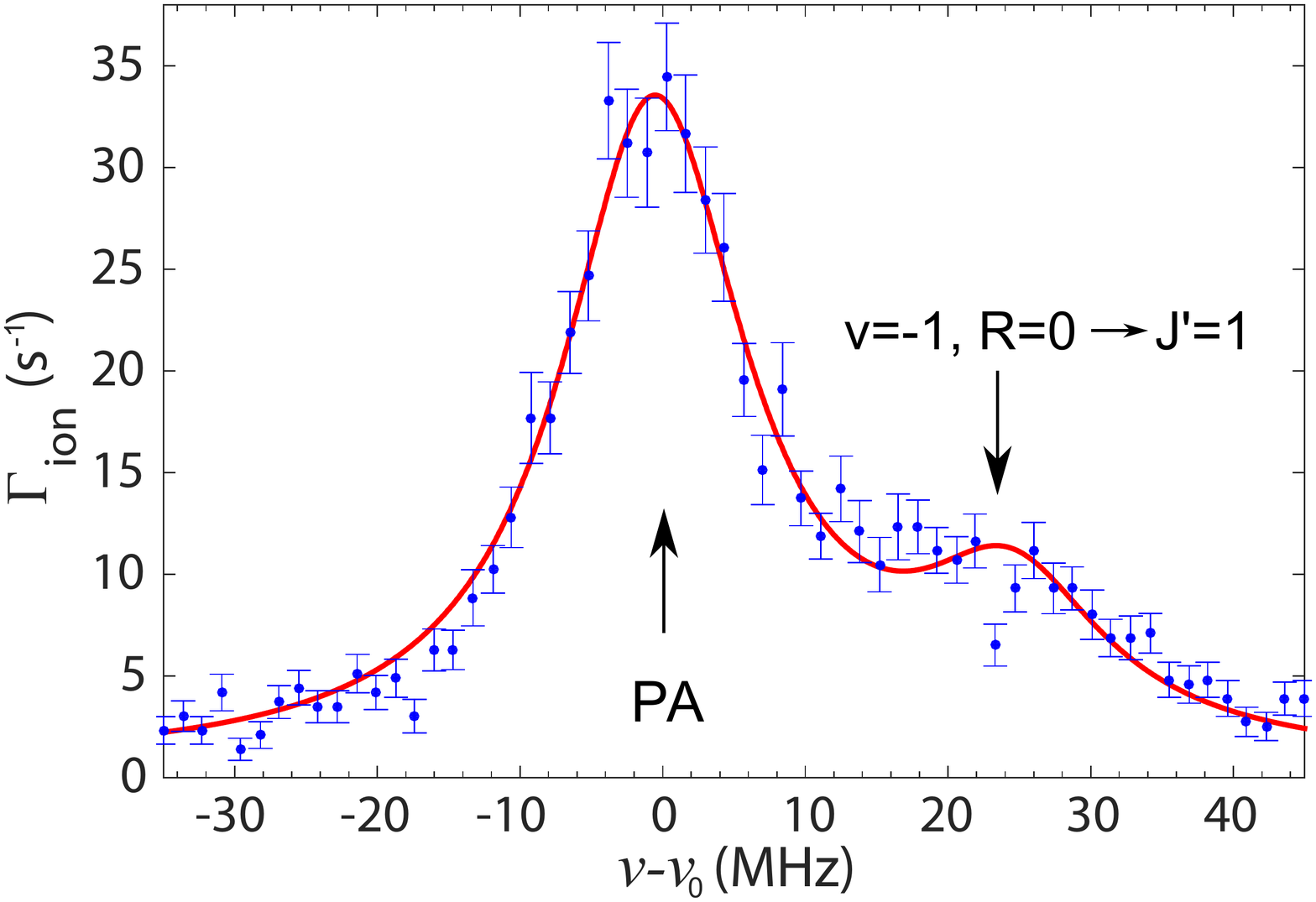}
	\end{center}
			\renewcommand{\figurename}{\textbf{Fig. 4:}}
	\renewcommand{\thefigure}{\hspace*{-2.2 mm}}
	\renewcommand{\labelsep}{none}
	\caption{\textbf{Detection of the most-weakly bound molecular state.}  Shown is a measured REMPI spectrum as a function of the probe laser frequency $\nu$. Next to the photoassociation line (PA) at $\nu = \nu_0 \equiv 281,445.045\:\textrm{GHz}$, there is a second peak at $\nu- \nu_0 = 24\:\textrm{MHz}$ which is the molecular product signal of the most-weakly bound state with even parity ($\textrm{v}=-1,R=0$).
 Each data point corresponds to 60 repetitions, and the error bars represent the statistical standard deviations.
  The red solid line is a fit of two Lorentzians to the data.}
	\label{fig:Fig3}
\end{figure}
\end{section}

\begin{section}*{Final-state population distribution}
We calibrate the ion signals in order to determine the three-body recombination induced loss rate constants $L_3(\textrm{v},R)$ for the  flux into individual rovibrational levels in absolute terms. For this, we determine the probability that a molecule is ionized by the REMPI scheme once it has been formed (see \cite{Supp}). Figure 5 (circles) shows the
individual loss rate constants $L_{3}(\textrm{v},R)$ as inferred from our experimental data (see \cite{Supp}).
 This corresponds to the population distribution of molecular products.
We investigate weakly bound molecular levels up to a binding energy of $17\:\textrm{GHz}\times h$ as marked by vibrational and rotational quantum numbers in the plot. The observed lines can unambiguously be assigned to molecular product states as measurements and calculated resonance frequencies accurately match within the experimental resolution of a few megahertz. Generally, the error bars shown in Fig.$\:$5 represent the statistical uncertainties of the ion signals. We note, however, that additional large uncertainties arise (see \cite{Supp}).

\begin{figure*}[t!]
	\begin{center}
		\includegraphics[width = 0.99\textwidth]{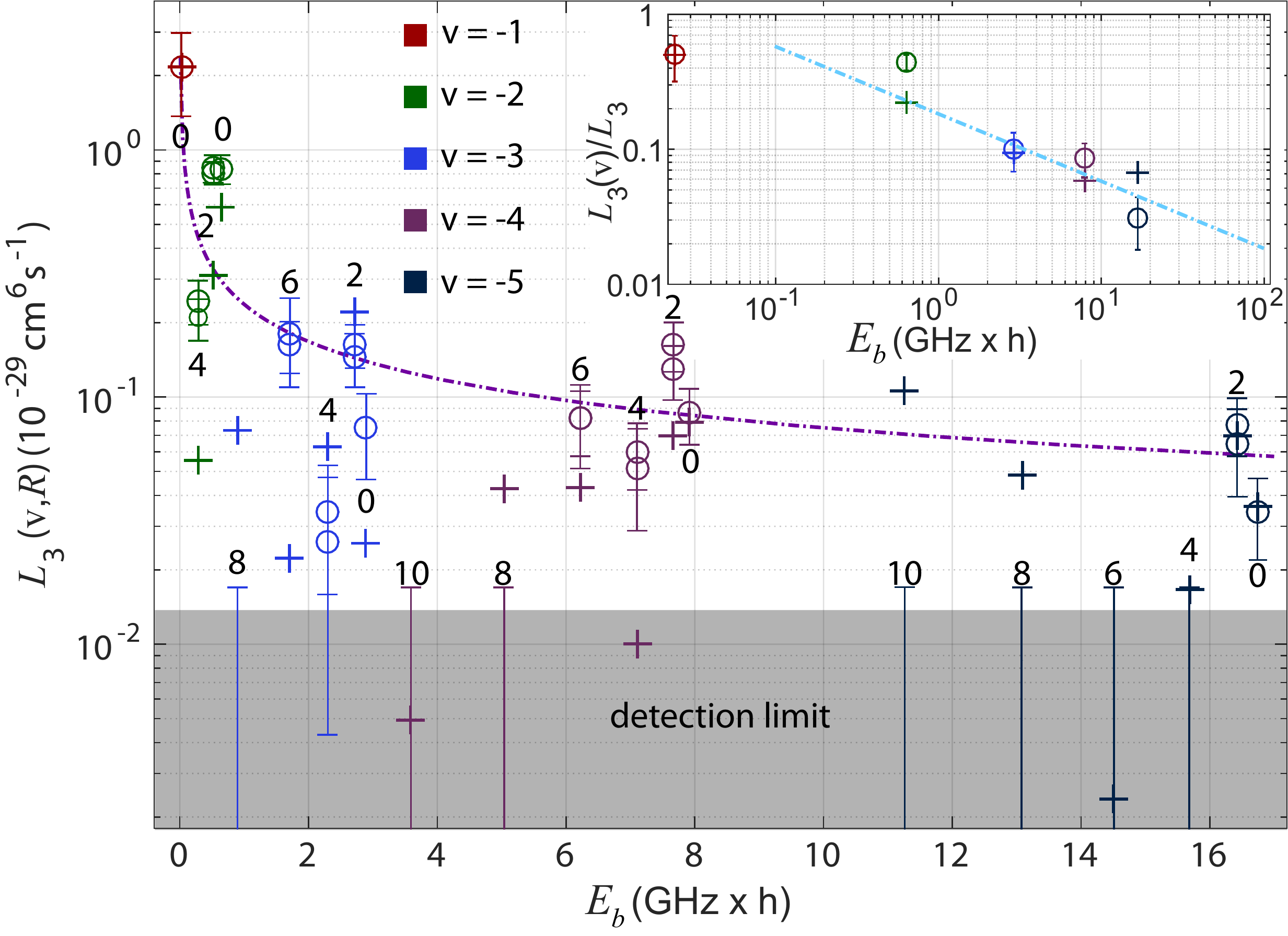}
	\end{center}
			\renewcommand{\figurename}{\textbf{Fig. 5:}}
	\renewcommand{\thefigure}{\hspace*{-2.5 mm}}
	\renewcommand{\labelsep}{none}
	\caption{\textbf{Population distribution of molecular product states following three-body recombination.}
The plot shows the loss rate constants $L_3(\textrm{v},R)$ due to three-body recombination into various molecular product channels, as specified by the quantum numbers $\textrm{v}$, $R$ and the respective binding energy $E_b$. $R$ is indicated next to the data points. All product channels belong to the $f_a=1, f_b=1 $ atomic asymptote and have $F = 2$. Circles are measurements. Crosses are calculations, rescaled as described in \cite{Supp}. Error bars in the grey region indicate upper limits derived from the experimental noise level. Two circles for the same product channel correspond to REMPI transitions to two different excited levels, $J' = R \pm 1$.   The inset presents the branching ratio into the five vibrational levels, calculated by summing over all respective rotational contributions and by normalizing with the total loss rate constant $L_3$. The dashed dotted lines (figure and inset) are proportional to  $1/ \sqrt{E_b}$ where $E_b$ is the binding energy. The error bars correspond to the statistical standard deviation.}
	\label{fig:Fig5}
\end{figure*}

Our data shows an overall tendency for the loss rate constants $L_3(\textrm{v},R)$ to drop for increasing binding energies as anticipated for a dilute ultracold gas, but with fluctuations with $R$ as discussed below. The decrease is not abrupt and roughly agrees with a $1/\sqrt{E_b}$ dependence, where $E_b$ is the binding energy.
This might indicate that $L_3(\textrm{v},R)$ is mainly determined by the time scale for the 3-atom collision complex to separate into its atomic and diatomic products, as this time scale is set by the inverse exit channel velocity.
The $1/\sqrt{E_b}$ dependence is  a simple propensity rule that can be tested further in future research.
 For understanding the measured population distribution, we have also carried out state-of-the-art numerical three-body calculations (crosses in Fig.$\:$5) based on a simplified model of long-range potentials (see \cite{Supp}). In agreement with the data, we find a general $1/\sqrt{E_b}$ dependence for $L_3(\textrm{v},R)$.

 For a fixed  $\text{v}$, both theory and experiment show large variations of $L_3(\textrm{v},R)$ with $R$.
 The rotational population distribution seems to mildly follow an overall pattern when comparing various vibrational levels.
 For example, the $R = 4$ signals seem to be generally suppressed with respect to the $R = 2$ signals.
 Our calculation  suggests that the variation of population with $R$ is of oscillatory nature.
  In general, variations as a function of $R$ are expected since the state-to-state $S$-matrix elements will be influenced by multiple paths and St\"{u}ckelberg oscillations associated with one or more curve crossings in the adiabatic potential curves, as indicated in \cite{Wang2011b,Esry1999,Incao2005} (see also Fig.$\:$S5 in \cite{Supp}).
 While in the experiment we  observe rotational angular momenta up to $R = 6$, theory predicts population of even higher rotational quantum numbers.
 In our calculations the variations for the $\text{v} = -2$ and $-3$ levels seem to have converged with respect to increasing the number $N_s$ of $R =0$ bound states in the model (see \cite{Supp}). However, the $\text{v}=-4$ and $-5$ rotational distributions have not yet fully converged with respect to adding bound states. Thus, the rotational distribution for these more-deeply bound levels is sensitive to shorter-range details that our model is not treating fully. Despite these limitations the theoretical model suggests that the rotational distributions for given $\text{v}$ might have non-negligible sensitivity to three-body corrections that are not pairwise additive (see \cite{Supp}).
 Our calculation also suggests that the oscillatory dependence on $R$ extends possibly to more-deeply bound levels.

In general, the sum $L_{3}(\textrm{v})=\sum\limits_{R}L_{3}(\textrm{v},R)$ over all rotational contributions for a given $\text{v}$ might be less subject to variation.
 The inset in Fig.$\:$5 presents the measured and calculated values of $L_{3}(\textrm{v})$ for the individual vibrational levels, normalized to the total loss rate of $L_3 = (4.3\pm 1.8) \times 10^{-29}\:\text{cm}^6\,\text{s}^{-1}$ determined by Burt {\em et al.} \cite{Burt1997}. Indeed, the results show a similar qualitative drop off tracking near a line that varies as $1/\sqrt{E_b}$. According to the given normalization the measurements reveal that roughly 50\% of all molecules produced via three-body recombination are formed in level $\text{v}=-1$.
We can estimate that about 10\% of the molecules are more deeply bound than $\text{v}=-5$, using the $1/ \sqrt{E_{b}}$ scaling law for the vibrational population.
 The total population of levels $\text{v} < -5$ is then approximately given by $ \sum\limits_{i < -5} E_b(v)^{-1/2} / \sum\limits_{v} E_b(v)^{-1/2} $. Here,  $E_b(v)$ is the  calculated $R = 0$ bound state energy for the vibrational level $\text{v}$ of the mixed $a^3\Sigma_u^+$ state. Indeed, some of the more deeply bound levels were observed in \cite{Haerter2013}.

\end{section}

\begin{section}*{Outlook}
In the near future we plan to extend the studies of the product population to more-deeply bound states to study how propensity rules change with binding energy.  Furthermore, it will be important to investigate how the product distribution depends on the initial quantum states and the scattering length of the colliding particles. Our experimental scheme can readily be adapted to other bosonic or fermionic elements or isotopes. In general, the product measurement technique introduced here can be used to investigate a wide range of  inelastic processes at  ultra-low temperatures well beyond three-body recombination such as molecular relaxation and rearrangement reactions. Therefore, the present work sets the stage for  experiments where chemical reactions  and inelastic collisions can be explored in a state-to-state fashion with full resolution on the quantum level.\\
Our comparison of measurement and state-of-the-art theoretical calculations demonstrates areas of both agreement and difference and should stimulate new theoretical efforts for pushing forward the current limits in the description of few-body dynamics. The theoretical considerations presented here suggest that simple models can help to deduce qualitative trends, but it is clear that much work on interaction potentials and numerical methods will be needed to develop a comprehensive understanding of recombination pathways and propensity rules. Our work discloses several experimental and theoretical directions to examine in future research. Besides gaining a better understanding, this work leads the path to new tools for controlling chemical reaction processes.

\end{section}

\newpage

\begin{section}*{Supplementary Materials and Methods}
\vspace{-0.4cm}
\subsection*{1. Preparation of the atomic sample and setting the total three-body loss rate}
\vspace{-0.4cm}
The atomic cloud is prepared in a crossed optical dipole trap at a wavelength of $1,064.5\:\textrm{nm}$ with trapping frequencies of $\omega_{x,y,z}= 2\pi\times (23,180,178)\:\textrm{Hz}$, where $z$ denotes the vertical direction.  The magnetic field is about $3\:\textrm{G}$. Typically, we work with a thermal sample at an initial temperature of about $750\:\textrm{nK}$ that consists of $N_\mathrm{at}\approx 5\times 10^6$ atoms. It is a Gaussian-shaped cloud and has a cloud size of $\sigma_{x,y,z}\approx(58.6,7.5,7.5)\:\upmu\textrm{m}$.
The initial peak particle density is $n_0\approx 0.9 \times 10^{14}\:\textrm{cm}^{-3}$. The atomic density distribution $n$ sets the total three-body loss rate $\dot{N}_\mathrm{at} =\int  \dot{n}\ d^3r =  -L_3  \int  n^3 d^3 r $, where $L_3$ is the total loss rate constant.  $L_3$ has been measured, e.g., by Burt \textit{et al.} \cite{Burt1997} to be $L_3=(4.3\pm1.8)\times 10^{-29}\:\textrm{cm}^6\,\textrm{s}^{-1}$.  Using this loss rate constant, $N_\mathrm{at}\approx 150,000$ atoms are typically lost due to three-body recombination in the first 500 ms.

\subsection*{2. Properties of molecular states and selection rules}
\vspace{-0.4cm}

In order to label the molecular quantum states of the weakly bound $a/X$-state molecules we use the atomic pair basis (Hund's case e), $|f_a, f_b, F, R, F_{tot}, m_{Ftot} \rangle$.
The angular momenta $\bm{F}$ and $\boldsymbol{R}$  couple to each other to form the total angular momentum  ${\bm{F}}_{tot}$, giving rise to level splittings for different $F_{tot}$ that are smaller than 1 MHz. In the present experiments we do not resolve this $F_{tot}, m_{Ftot}$ substructure.
 According to the selection rules for optical dipole transitions, only molecules containing a $X^1\Sigma_g^+$ component can be excited towards $A^1\Sigma_u^+$. Fortunately, a majority of the weakly bound molecular states exhibit a sizeable singlet admixture of at least 10\%, due to the hyperfine interaction (see Supplementary Data). For Rb$_2$ these states are characterized by $F = 2, 0$ ($F = 1, 3$) for positive (negative) total parity, respectively.

The excited state $A^1\Sigma_u^+$, $\textrm{v}'=66$  has hyperfine splittings off less than 3$\:$MHz, despite the fact that spin-orbit coupling mixes in 16\% of the $b^3\Pi_u$ state (see \cite{Drozdova2013,Deiss2015}).  Therefore $A^1\Sigma_u^+$, $\textrm{v}'=66$  exhibits an essentially pure rotational substructure $\propto B_\text{v}' \, J'(J'+1)$  with a rotational constant of $B_\text{v}' = 443\:$MHz.
For exciting a molecule towards the state $A^1\Sigma_u^+$, optical dipole selection rules demand $R \rightarrow J' = R + 1$ for $R = 0$, and $R \rightarrow J' = R \pm 1$, otherwise.
\vspace{-0.4cm}
\subsection*{3. Coupled-channel calculations}
\label{sec_CC}
\vspace{-0.4cm}
The Rb$_2$ energy levels are calculated with a coupled-channel model which uses the Hamiltonian for atom pairs of $s+s$ or $s+p$ states (see e.g. \cite{SStrauss2010,SDrews2017}) and potentials derived from former spectroscopic work of several groups. We have reevaluated the analysis of the singlet and triplet ground states \cite{SStrauss2010} correcting the assignment of some high rotational levels as remarked in \cite{SGuan2013} and derived improved potentials for $X^1\Sigma^+_g$ and $a^3\Sigma^+_u$.
  Because the analysis includes a number of Feshbach resonances we obtain reliable predictions for the asymptotic level structure. The potential system for the atom pair $s+p$ was derived from spectroscopic work collected in \cite{Drozdova2013} and work from our own group \cite{Deiss2015,SDrews2017}. The coupled-channel calculation describes in full detail the singlet-triplet mixing needed for deriving reliable Franck-Condon factors. It predicts the hyperfine structure in the state $A^1\Sigma^+_u$ to be small, such that it is negligible for the present analysis.

\vspace{-0.4cm}
\subsection*{4. REMPI lasers}
\vspace{-0.4cm}
The first excitation step for the REMPI is driven by a cw external-cavity diode laser (the \grq probe laser'). It has a Gaussian beam waist ($1/e^2$-radius) of $\approx280\:\upmu\textrm{m}$ at the position of the atomic sample, so that the intensity is $\approx12\:\textrm{W\,cm}^{-2}$. The second REMPI step is driven by the dipole trap laser (the \grq ionization laser'). Since we work with a crossed dipole trap, the ionization laser consists of two beams. The beam in the horizontal (vertical) direction is focused to a beam waist of $\approx90\:\upmu\textrm{m}$ ($\approx150\:\upmu\textrm{m}$) with an intensity of $\approx34\:\textrm{kW\,cm}^{-2}$ ($\approx8\:\textrm{kW\,cm}^{-2}$), and their relative detuning is always $160\:\text{MHz}$. The wavelength of the ionization laser is fine tuned for producing a minimal  background ionization signal of $\approx 1\:\textrm{s}^{-1}$ when the probe laser is turned off. The probe and ionization lasers have a short-term frequency stability on the order of $100\:\textrm{kHz}$ and $1\:\textrm{kHz}$, respectively. Both are stabilized to a wavelength meter achieving a shot-to-shot and long-term stability of a few megahertz. The polarization of the lasers can equally drive $\sigma$- and $\pi$-transitions.

\vspace{-0.4cm}
\subsection*{5. Ion detection}
\vspace{-0.4cm}
The Rb$_2^+$ ions produced via the REMPI process are first captured in a $1\:\textrm{eV}$ deep Paul trap before they are detected with near unit efficiency. During the process each Rb$_2^+$ molecule typically dissociates into $\text{Rb}^++ \text{Rb}$. However, the number of trapped ions is conserved. Since fluorescence detection is neither available for  Rb$^+$ nor for Rb$_2^+$, we use the following two detection methods.

\textit{Ion detection scheme I}:
The trapped ion cloud is centered on the cold atom cloud. Thus, from the moment that the ions are produced, they can elastically collide with the neutral atoms and expel them from the shallow dipole trap. Typically, after $500\:\textrm{ms}$ we end the experimental run, and measure the number of remaining atoms. The larger the atom loss, the larger the number of trapped ions. Detection scheme I is fast, allowing for a high repetition rate. However, it is not very precise in measuring the absolute number of trapped ions. For ion detection scheme I we operate the Paul trap with a micromotion energy of about $1\:\textrm{K}\times k_\mathrm{B}$, where $k_\mathrm{B}$ is Boltzmann's constant. At these energies, three-body recombination of two Rb atoms with an ion is strongly suppressed \cite{Kruekow2016}, and two-body collisions of an atom and an ion dominate.

\textit{Ion detection scheme II}: Here, the centers of the optical dipole trap and of the Paul trap are separated by about $300\:\upmu\textrm{m}$ from each other, similar to the configuration given in H\"{a}rter \textit{et al.} \cite{Haerter2013}. Consequently, a produced Rb$_2^+$-ion is quickly pulled outside of the atom cloud into the ion trap such that further atom-ion collisions are suppressed. In order to trap one to three ions we typically set an appropriate accumulation time in the range between $2$ to $500\:\textrm{ms}$. We count the trapped ions by immersing them into a new atom cloud with a comparably low density of $4\times 10^{12}\:\textrm{cm}^{-3}$ for an interaction time of $1.5\:\text{s}$. As in detection scheme I, the ions inflict atom losses which increase stepwise with the number of ions, see \cite{Haerter2012,Haerter2013b}. The scheme can count up to five ions. The micromotion energy is set to about $0.1\:\textrm{K}\times k_\mathrm{B}$.

\vspace{-0.4cm}
\subsection*{6. Density dependence of ion production rate}
\vspace{-0.4cm}

In order to check whether resonance lines of our REMPI spectra are the result of three-body recombination we measure how the ion production rate depends on the atomic density.
 Figure $\:$S1 shows the ion signal for the level $\text{v} = -2, R = 0$ after an accumulation time of 100 ms. The measured signal has been normalized to represent the ion signal per million atoms in the cloud. The scaling in terms of the atomic density $n$ is clearly quadratic, as expected for a three-body process. A two-body process such as photoassociation would be characterized by a linear dependence.

\begin{figure*}[h!]
	\begin{center}
		\includegraphics[width = 0.7\textwidth]{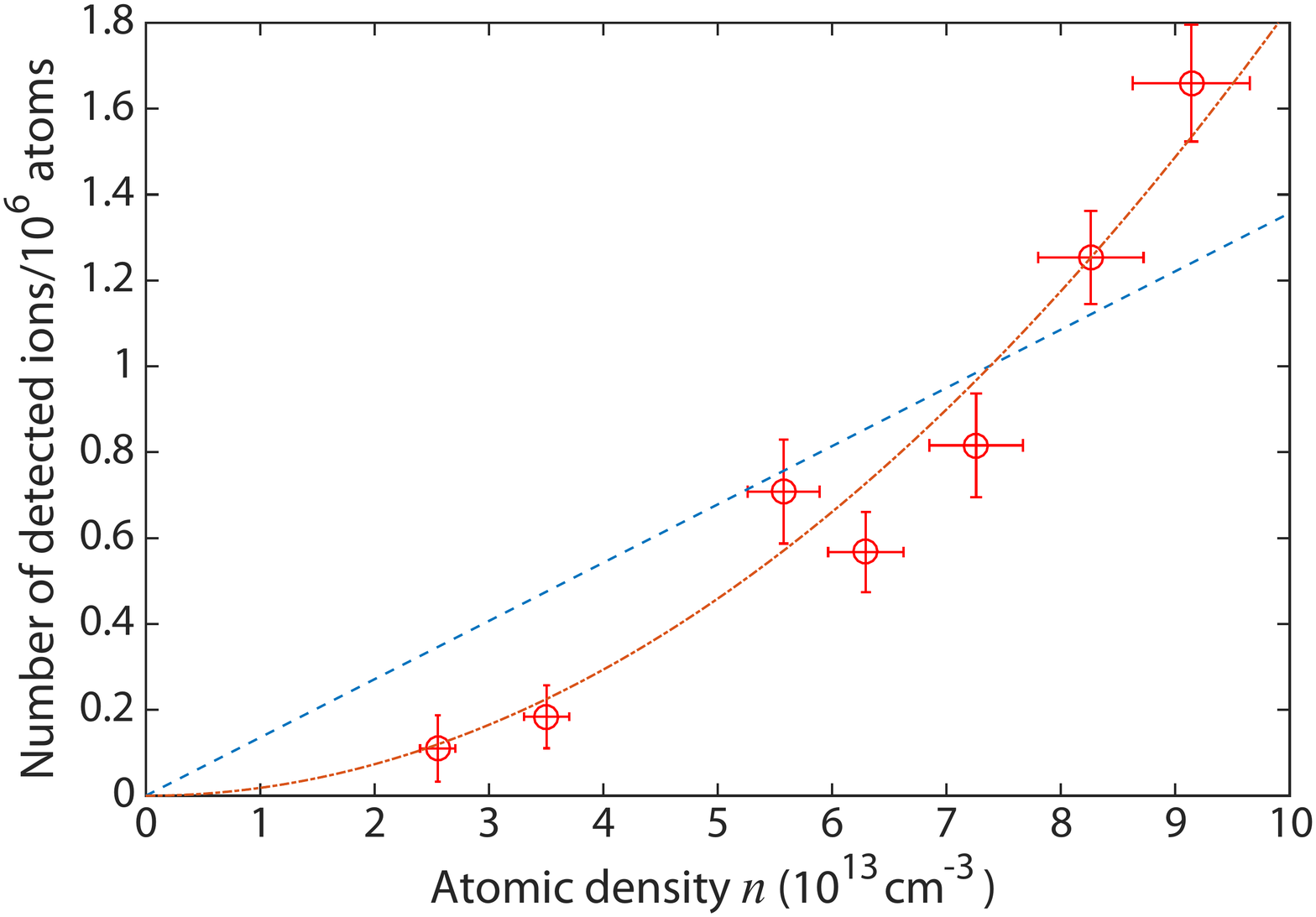}
	\end{center}
	\renewcommand{\figurename}{\textbf{Fig. S1:}}
	\renewcommand{\thefigure}{\hspace*{-2mm}}
	\renewcommand{\labelsep}{none}
	\caption{\textbf{Dependence of the ion production rate on atomic density.} The ion signal has been  normalized by the atom number to represent the ion signal per million atoms in the cloud. The blue dashed line corresponds to a linear fit and the orange dash-dotted line to a quadratic fit.}
	\label{Supfig:Fig1}
\end{figure*}

\vspace{-0.4cm}
\subsection*{7. Detection efficiency as a function of probe laser intensity}
\vspace{-0.4cm}

We experimentally investigate the ionization signal of $\textrm{v}=-2$ molecules as a function of the probe laser power, see Fig.$\:$S2. For powers of more than 15$\:$mW, the signal is strongly saturated for all rotational states. The continuous lines are fits of the model function $S(P)= S_{\text{max}}(1-\exp(-C \cdot P))$ to the data, where $P$ is the power of the probe laser and $C$ is a fit constant.
  We have verified that this model function describes well the results of our Monte-Carlo calculations for the expected ionization signal (see Section 8). In our experiments for $\textrm{v}=-2$ and more-deeply bound states we typically work with a power of 30$\:$mW of the probe laser beam. Using simple scaling laws and taking into account the respective Franck-Condon and H\"{o}nl-London factors, the singlet content of the total electronic spin state and the Doppler shifts of the molecules, we have checked that also the signals of the more-deeply bound levels should be strongly saturated.
  Furthermore, we observe that the ion signals for the transitions $R \rightarrow J' = R \pm 1$ are of similar strength, see, e.g. Fig.$\:$2 in the main text.  (The disparity of the $R = 2 \rightarrow J' = 1, 3$ curves in Fig.$\:$S2 is probably a result of measuring the $R = 2 \rightarrow J' = 1$ transition slightly off-resonance.) In general, our measurements suggest that the signal saturation is  maintained for rotational states up to $J' = 7$.

\begin{figure*}[h!]
	\begin{center}
		\includegraphics[width = 0.7\textwidth]{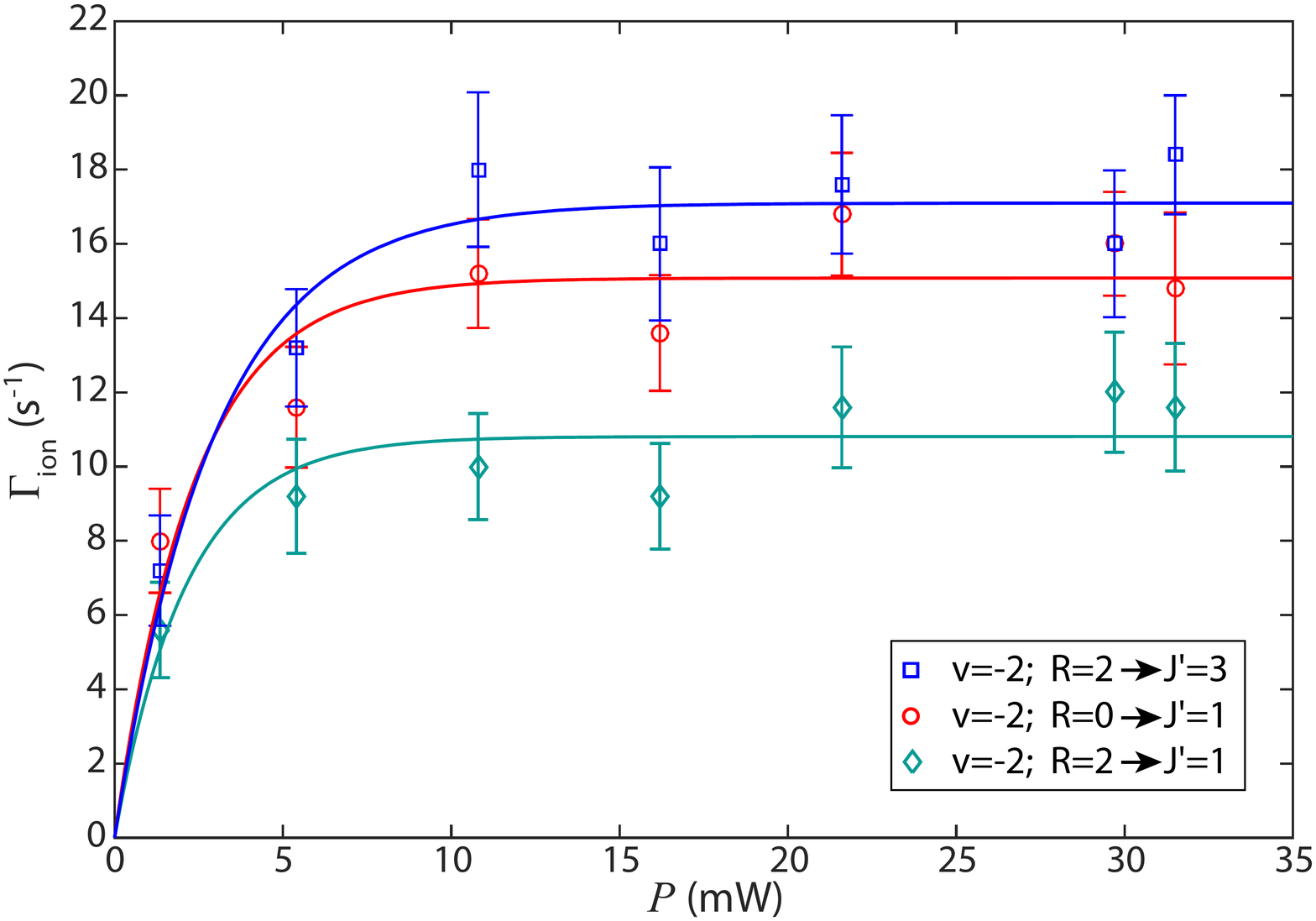}
	\end{center}
	\renewcommand{\figurename}{\textbf{Fig. S2:}}
	\renewcommand{\thefigure}{\hspace*{-2.2 mm}}
	\renewcommand{\labelsep}{none}
	\caption{\textbf{Detection signals of two rotational levels of $\textrm{v}=-2$ molecules as a function of the probe laser power}. We investigate REMPI signals for three different $X^1\Sigma_g^+ \rightarrow A^1\Sigma_u^+$ transitions as indicated by the legend. The continuous lines are fits of the model function $S(P)= S_{\text{max}}(1-\exp(-C \cdot P))$ to the data. The fit constant is roughly the same $C\approx 2.5\:\text{(mW)}^{-1}$ for all three curves. $P$ is the total power of the probe laser.
	}
	\label{Supfig:Fig2}
\end{figure*}

\vspace{-0.4cm}
\subsection*{8. Monte-Carlo model calculations}
\label{secMonteCarlo}
\vspace{-0.4cm}
Once a weakly bound Rb$_2$ molecule is formed via three-body recombination it will fly out of the atomic cloud with a velocity $\text{v}_{\text{Rb2}} = \sqrt{ E_b/(3m_{\text{Rb}})}$ which is determined by the released binding energy $E_b$ and the mass of the rubidium atom $m_{\text{Rb}}$. On the way out, the molecule can  collide with cold atoms and relax to a more-deeply bound state. The rate for this process is
$\Gamma_{\text{rel}} = K_{\text{rel}}\, n({\bm{r}})$, where $n({\bm{r}})$ is the local atomic density and $K_{\text{rel}}$ is the relaxation constant. We assume $K_{\text{rel}} = 10^{-10}\:\text{cm}^3\,\text{s}^{-1}$, see \cite{SMukaiyama2004,SStaanum2006,SZahzam2006,SQuemener2007,SQuemener2005}. At the same time, the probe laser can photoexcite the molecule with a rate $\Gamma_{\textrm{phot}} = \tilde{c} \  I({\bm{r}}) \ f_{\text{FC}} \ f_{\text{HL}} \ / ( (\gamma / 2)^2 + \delta^2)$, where $I({\bm{r}})$ is the local intensity of the Gaussian laser beam, $f_{\text{FC}}$ is the Franck-Condon factor for the singlet parts of the multicomponent molecular wave functions,
$f_{\text{HL}}$ is the H\"{o}nl-London factor,
 \begin{figure*}[b]
 	\begin{center}
 		\includegraphics[width = 0.9\textwidth]{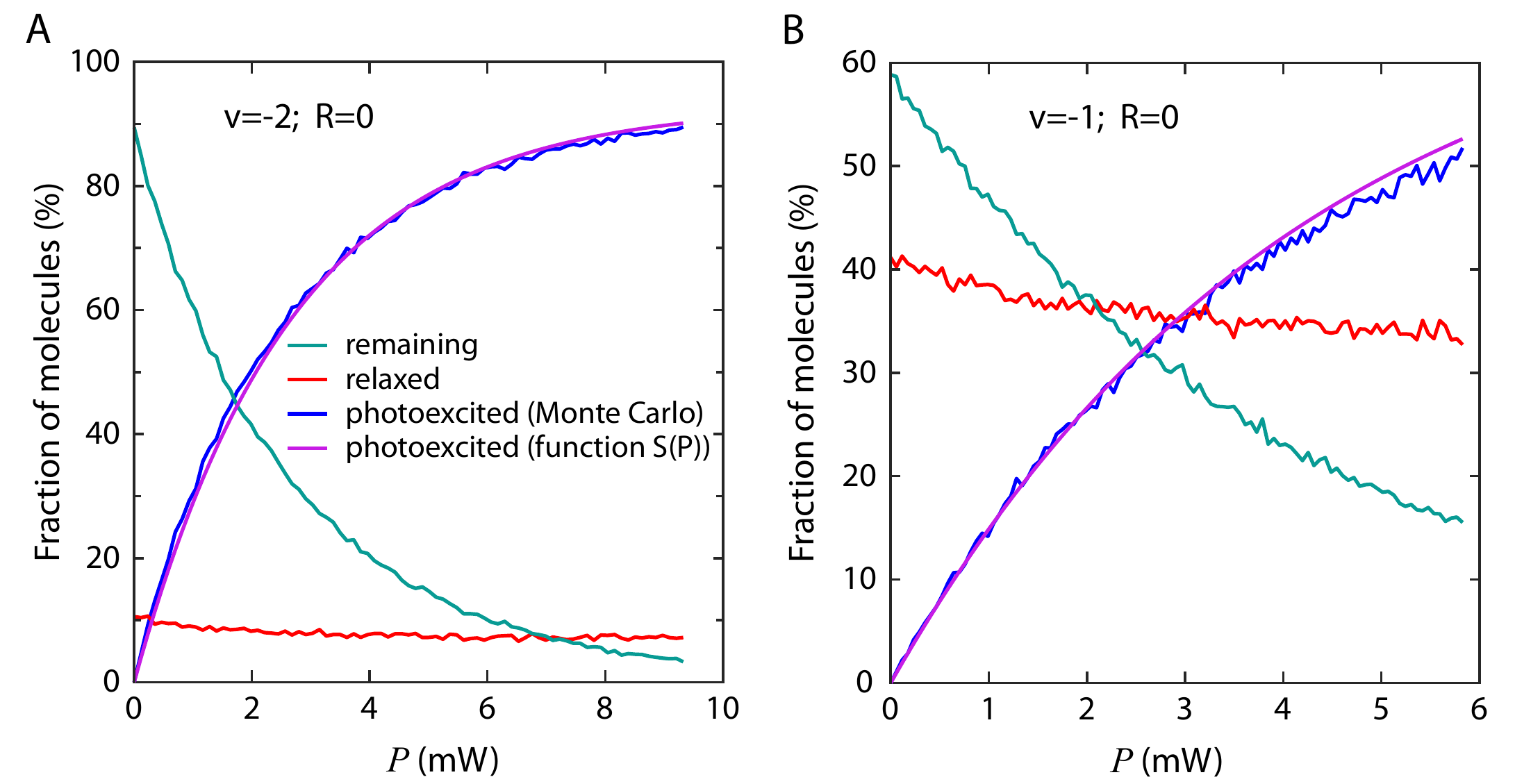}
 	\end{center}
 	\renewcommand{\figurename}{\textbf{Fig. S3:}}
 	\renewcommand{\thefigure}{\hspace*{-2mm}}
 	\renewcommand{\labelsep}{none}
 	\caption{\textbf{Monte-Carlo model calculations.}   The fractions of molecules that remain unscathed (turquoise), that relax to more-deeply bound states due to collisions with atoms (red), or that are photoexcited (blue), are plotted as a function of the probe laser power. \textbf{A} and \textbf{B} show the results for $\textrm{v}=-2,-1$ molecules, respectively.  The smooth continuous lines (purple) have the form $\tilde{S}(P)=\tilde{S}_{max}(1 - \exp(- C \cdot P)) $, where $P$ is the total probe laser power and   $\tilde{S}_{max}$ as well as $C$ can be adapted to achieve a reasonable fit.}
 	\label{Supfig:Fig3}
 \end{figure*}
 $\gamma \approx 10\:\text{MHz}$ is the linewidth of the excited molecular level, $\delta$ is the laser detuning from resonance, and $\tilde{c}$ is a constant that includes, e.g., the transition electric dipole matrix element. We calculate Monte-Carlo trajectories of produced molecules and determine the probability that the molecules undergo relaxation or photoexcitation within the transient time. The molecules are created according to the spatial probability distribution $ n^3({\bm{r}}) / \int n^3({\bm{r}}) d^3 r$  with the direction of their respective velocity being random. On average, the molecular velocities will lead to Doppler-broadening and to a reduction of the on-resonance photoexcitation rate by a factor given by $\arctan ( 2 \textrm{v}_{\text{Rb2}} / (\lambda \gamma ))  / (2 \textrm{v}_{\text{Rb2}} / (\lambda \gamma))$,  where $\lambda$ is the transition wavelength. However, this reduction can be neglected for our purposes.
 
Figure S3 A  shows the Monte-Carlo calculations for the $\textrm{v}=-2$ level for the experimental parameters of our setup.
The presented curves are the probabilities that for a given power a created molecule is photoexcited, undergoes relaxation, or remains unscathed.
For example, at a resonant probe laser power of $9\:\text{mW}$ about 89\% of the produced $\text{v}=-2$ molecules will be photoexcited. About 8\% will be lost due to relaxation collisions and about 3\% will leave the cloud and probe laser beam without having relaxed or being photoexcited.
We find that the photoexcited fraction can be well described by the model function $\tilde{S}(P)=\tilde{S}_{max}(1 - \exp(- C \cdot P)) $.
In order to reproduce the power dependence of the $\textrm{v}=-2$ data in Fig.$\:$S2 we have adjusted $\tilde{c}$, which effectively calibrates the total power $P$ in our calculations.
Keeping the same $\tilde{c}$, the Monte-Carlo model can now also be used for any other vibrational level, if  the respective values for
$f_{\text{FC}}$, $f_{\text{HL}}$, and $\text{v}_\text{Rb2}$ are employed. In Fig.$\:$S3 B we show calculations for $\text{v}=-1$.

\vspace{-0.4cm}	
\subsection*{9. Calibration of REMPI efficiency}
\vspace{-0.4cm}
The overall efficiency of REMPI is the product of the efficiency $\eta_1$ to excite the molecule (within its lifetime in the REMPI laser beams) to the  $A^1\Sigma_u^+, \text{v}'=66$ level and the efficiency $\eta_2$ to ionize this excited molecule.

In order to determine $\eta_1$, we use both our measurements and calculations presented in the Sections 7 and 8.
With the exception of level $\text{v}=-1$, we generally work with probe laser powers of about $30\:\textrm{mW}$ for which the ion signal is strongly saturated (see e.g. Fig.$\:$S2) and the observed linewidths of REMPI signals exceed the natural linewidth of the $A^1\Sigma_u^+$ level ($\approx10\:\textrm{MHz}$) by roughly a factor of two (see e.g. Fig.$\:$3). Despite an apparent saturation of the ion signal it is still possible that a fraction of molecules escape or relax in a collision before they are optically excited. Using our Monte-Carlo model we have calculated this fraction (see Fig.$\:$S3 A). We find that for all the vibrational levels in our measurements with $\text{v}<-1$, the excitation efficiency is $\eta_1>0.9$, for a probe laser power of $\approx 30\:\text{mW}$. In contrast, for the $\text{v}=-1$ level we work in a low power regime where the ion signals are not saturated. If, e.g., we work with a probe laser power of $1.5\:\text{mW}$, Fig.$\:$S3 B shows that only  21\% of the initially created $\textrm{v}=-1$ molecules are excited in the first step of our REMPI detection scheme, thus we derive $\eta_1 = 0.21$.

We determine the efficiency $\eta_2$ with the help of photoassociation. From a cold  Rb atom cloud we photoassociate several percent of the atoms to the intermediate level $A^1\Sigma_u^+$, $\textrm{v}'=66$, $J'=1$ within $1\:\text{s}$.  We deduce the number of photoassociated molecules by measuring the corresponding atomic losses from the atomic cloud via absorption imaging. 
We find that a small fraction $\eta_2$ of these molecules is ionized by the ionization laser. Measuring the number of  produced ions with detection scheme II, we determine the ionization efficiency to be $\eta_2=(6.6\pm0.7 \pm 2.2)\times 10^{-4}$. Here, the first uncertainty value is of statistical nature, while the second one mainly reflects systematic errors in the atom number measurements by absorption imaging.
We note that photoassociated molecules that are not ionized, relax quickly to the molecular ground state $X^1\Sigma_g^+$ by emitting a photon. At this point they are cold and trapped in the atomic cloud, but in general vibrationally excited. Within a few tens of ms, these molecules will then inelastically collide with another atom and relax to more deeply-bound vibrational states. The released binding energy will kick both the molecule and the atom out of the trap. Therefore each photoassociated molecule typically leads to a loss of three atoms from the cloud, a fact that needs to be taken into account when determining the number of photoassociated molecules from the measured atom loss.

\vspace{-0.4cm}	
\subsection*{10. Determination of molecule production rates}
\vspace{-0.4cm}	

 Once the REMPI efficiency $\eta_1\eta_2$ is determined, it is used to directly convert a measured ion number into a number of molecules $M(\textrm{v},R)$ formed in a particular state. The corresponding loss rate constant $L_3(\textrm{v},R)$ is calculated as
\begin{equation}
L_3(\textrm{v},R) =  {3\  M(\textrm{v},R) \over  \int_0^\tau  \int n({\bm{r}},t)^3 d^3 r\ dt},
\label{eq:2}
\end{equation}
 where $\tau$ is the accumulation time during which molecules are produced via three-body recombination. 
We determine the time dependent density $n({\bm{r}},t)$ by measuring the total atom number and temperature of the atom cloud at various times during its decay (see Fig.$\:$S4 in Section 11).

We note, that it is in general not a-priori clear whether all the molecules in a particular level originate directly from three-body recombination, as relaxation of more weakly bound molecules due to a collision with an atom might also contribute to the population of the level.
Clearly, the level $\text{v} = -1$ is most relevant for this process, and according to the Monte-Carlo calculations (see Section 8), about 40\% of the $\text{v}=-1$  molecules are expected to relax
for our given experimental parameters. Therefore,  as an example we tested whether the $\text{v}=-2$ level is strongly populated through relaxation of $\text{v} = -1$ molecules by repeating the recombination rate measurements for clouds with a factor of five smaller atom number and density.
Possible relaxation contributions to the flux into level $\text{v}=-2$ level should then be strongly suppressed. Our measurements, however, do not indicate such a suppression, as we obtain a similar loss rate constant $L_3(\textrm{v},R)$ as before. Therefore, this is evidence that the measured molecular fluxes are mainly originating from a direct three-body recombination into each $\textrm{v},R$ product channel.
Nevertheless, the determined population of the $\text{v}=-2,-3,-4, -5 $ levels has to be considered as an upper limit for three-body recombination flux.
\vspace{-0.4cm}
\subsection*{11. Decay of atomic cloud}
\vspace{-0.4cm}

We measure the decay of the number of Rb atoms trapped in our optical dipole trap while the probe laser is switched off (see Fig.$\:$S4). For times $t>5\:\text{s}$, the decay is exponential, mainly due to collisions with thermal background gas. For short times, additional losses, e.g., due to three-body recombination, photoassociation, and evaporation are present.

\begin{figure*}[h!]
	\begin{center}
		\includegraphics[width = 0.7\textwidth]{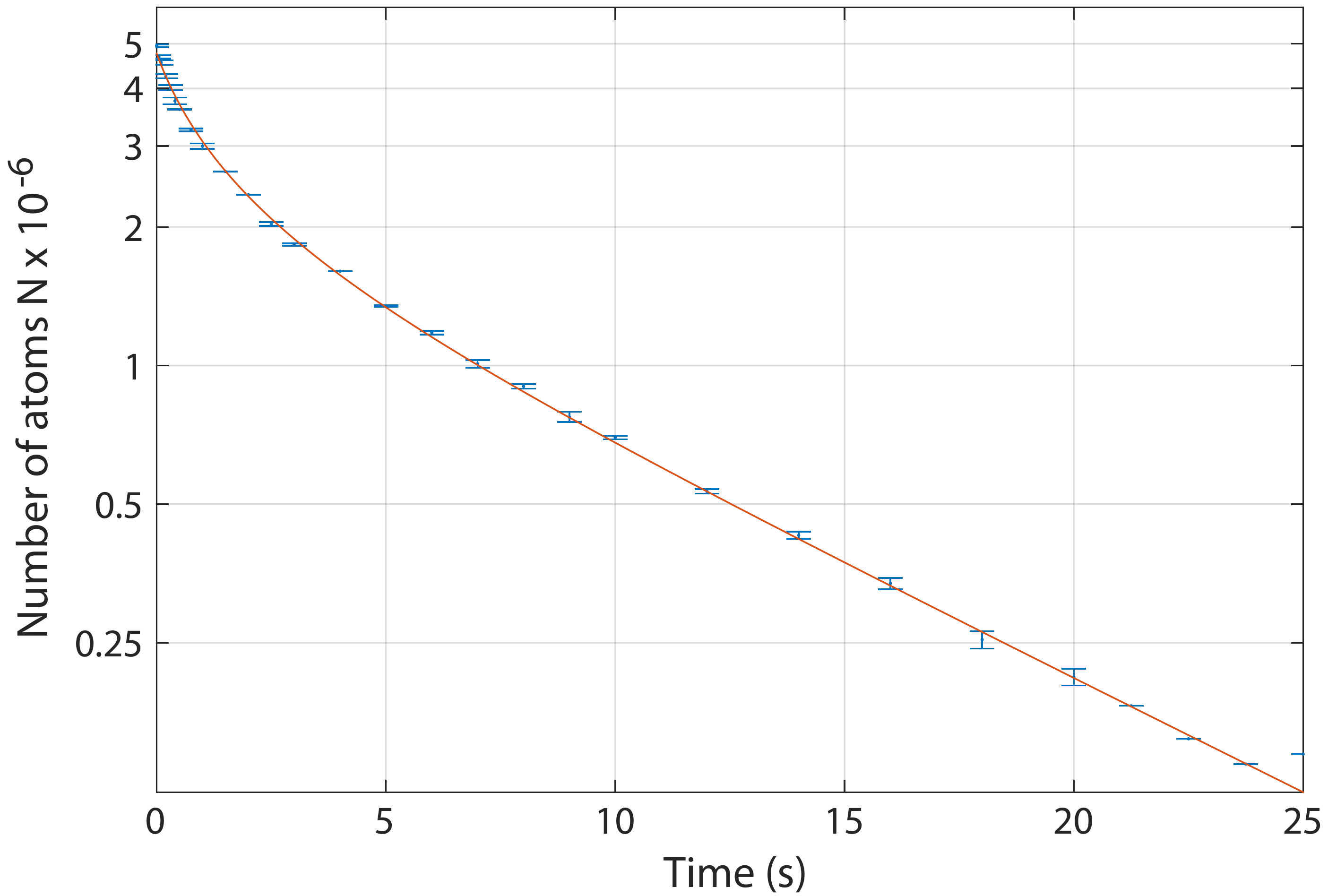}
	\end{center}
	\renewcommand{\figurename}{\textbf{Fig. S4:}}
	\renewcommand{\thefigure}{\hspace*{-2mm}}
	\renewcommand{\labelsep}{none}
	\caption{\textbf{Decay of the atomic cloud in the optical dipole trap.}
		 The number of remaining atoms of the atom cloud is plotted as a function of time. The continuous line is a guide to the eye. 	}
	\label{Supfig:Fig5}
\end{figure*}
\vspace{-0.4cm}	
\subsection*{12. Three-body model and calculations}
\vspace{-0.4cm}
Besides the experimental determination we calculate the set of loss rate constants $L_3(\textrm{v},R)$ numerically. Our computational method uses the adiabatic hyperspherical representation to solve the 3-body Schr\"{o}dinger equation \cite{Wang2011b,Mehta2009}. From the solution of the coupled equations, the elements of the unitary scattering $S$-matrix are obtained that connect the entrance channel state to the various product channel states $\textrm{v},R$. Consistent with the experimental observations, we ignore nuclear spin degrees of freedom so that we assume entrance three-body continuum states of total angular momentum ${\bm{J}}=0$  and positive parity. Consequently the product channels also have ${\bm{J}}=0$, resulting from the sum of the angular momenta for diatomic molecular rotation ${\bm{R}}$ and the relative angular momentum of the atom-diatom pair. The total three-body decay rate of an atomic gas of density $n$ is $\dot{n}=-L_3 n^3$, where $L_3=\sum_{\textrm{v},R}L_3(\textrm{v},R)$.

We follow Refs.$\:$\cite{Wang2014,Wang2012} in assuming that the three-body potential is a pairwise additive sum of two-body Lennard-Jones potentials $V(r)=-(C_6/r_s^6)(1-(\tilde{\lambda}/r_s)^6)$, where $r_s$ is the pair separation and $C_6=4710.22\:$a.u. is the known long-range van der Waals coefficient for two Rb atoms in their ground electronic state \cite{SStrauss2010}. The short-range parameter $\tilde{\lambda}$ is adjusted for a selected number $N_s$ of $R=0$ bound states to give the known scattering length $100.36\:a_0$ for two $^{87}$Rb atoms in their $f=1$, $m_f=-1$ hyperfine state \cite{SStrauss2010}. Here we let $N_s$ range from 1 to 6. For a similar example of calculations with Cs atoms see \cite{Wang2012}.

Figure S5 shows the three-body adiabatic potentials as functions of hyperspherical radius $\rho$ calculated for the case of $N_s=4$ $s$-wave bound states.  Three atoms come together from large $\rho$ with collision energy $E_{\textrm{col}}>0$  close to the $E_{\textrm{col}}=0$ threshold and encounter one another in the shorter range region of the potential, where they may recombine and separate to the product states of a dimer and a free atom.  The shared energy release in the separating products is $E_{\textrm{col}}-E_{b}(\textrm{v},R) \approx -E_{b}(\textrm{v},R)$.

\begin{figure*}[h!]
	\begin{center}
		\includegraphics[width = 0.8\textwidth]{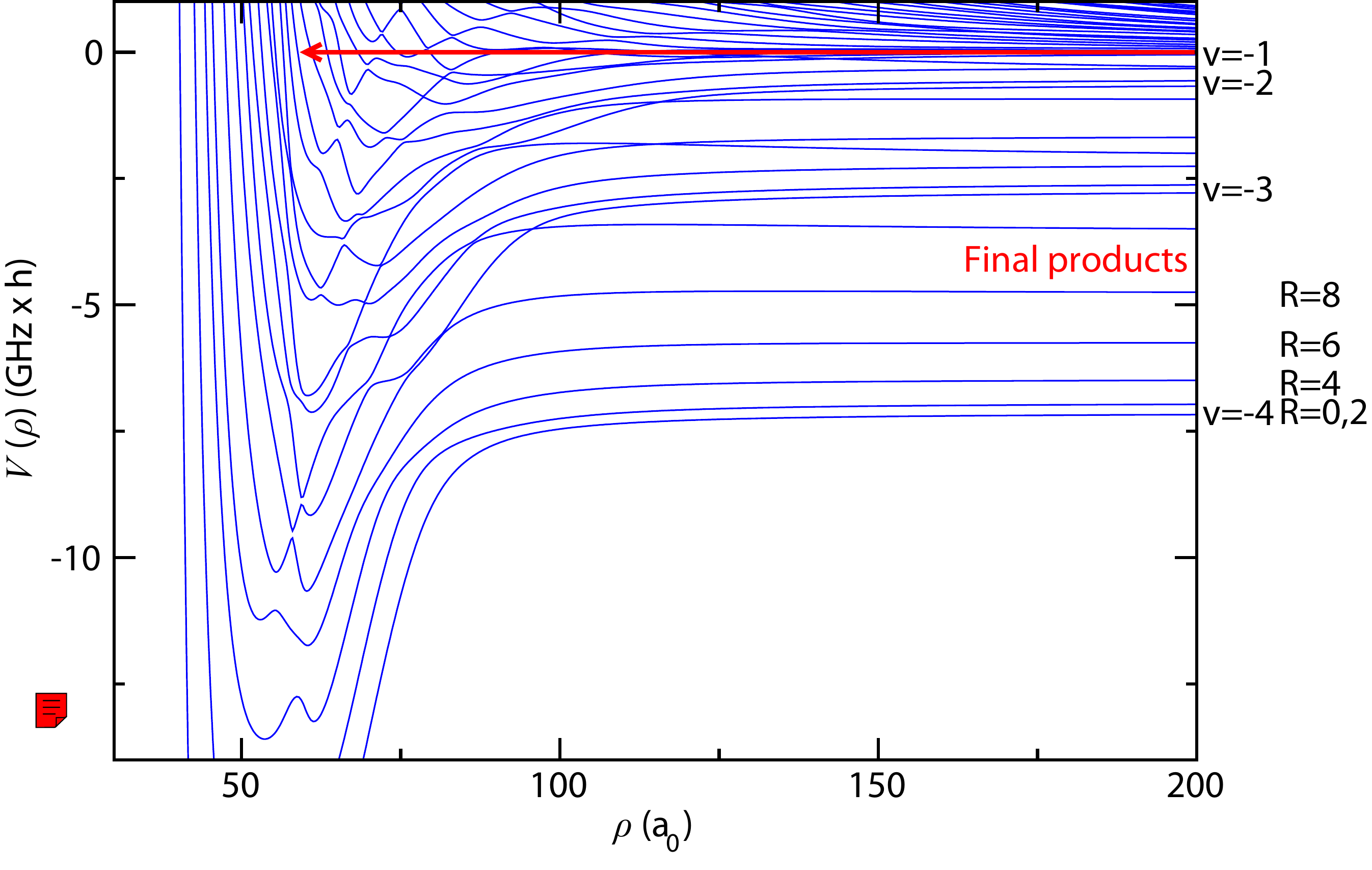}
	\end{center}
	\renewcommand{\figurename}{\textbf{Fig. S5:}}
	\renewcommand{\thefigure}{\hspace*{-2.5 mm}}
	\renewcommand{\labelsep}{none}
	\caption{ \textbf{Adiabatic hyperspherical potential energy curves $V(\rho)$ versus hyperspherical radius $\rho$  for three $^{87}$Rb atoms.}
		The number of $s$-wave bound states is $N_s= 4$; the vibrational and the rotational quantum number is shown for dimer product vibrational levels v $=$ $-1$, $-2$, $-3$, and $-4$.  The arrow indicates the three-body entrance channel threshold at energy $E_{\textrm{col}}=0$.  The potential energy curve asymptotes at negative energy
		correspond to dimer molecular levels.
		All states in the diagram have total angular momentum $J=0$. a$_0$ is the Bohr radius.}
	\label{Supfig:Fig65}
\end{figure*}

Figure S6 shows the calculated $L_3 (\textrm{v}, R)$ values for the case of $N_s = 6$. It is evident that the overall pattern exhibits two clear features: (1) a quite slow overall drop off with increasing binding energy, and (2) strong fluctuations with rotational quantum number $R$. The slow drop off has the form of $1/ \sqrt{E_b}$ and may be related to the classical time for the atom-diatom products to separate from the shorter range ``collision complex.'' The strong fluctuations show an oscillatory pattern and an overall decrease in $L_3$ with increasing $R$. This behavior is likely related to the multi-path interference in the shorter range region where the entrance and exit channels can be interconnected, reflected by the presence of multiple avoided crossings of the potentials in Fig.$\:$S5. As a result, there are many pathways on which phase interference might be operative, thereby leading to the fluctuations in the product distributions.

\begin{figure*}[h!]
	\begin{center}
		\includegraphics[width = 0.7\textwidth]{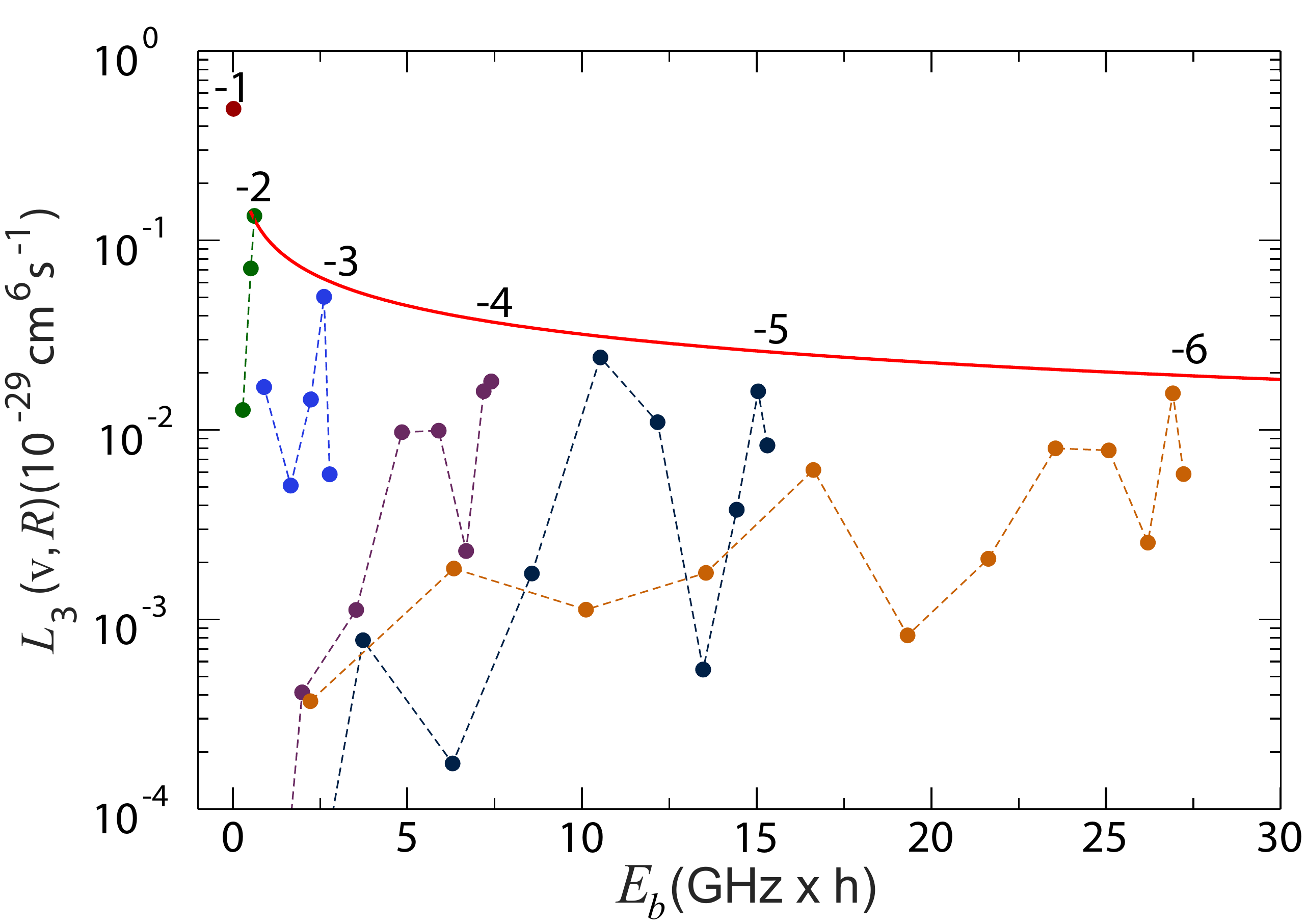}
	\end{center}
	\renewcommand{\figurename}{\textbf{Fig. S6:}}
	\renewcommand{\thefigure}{\hspace*{-2.1 mm}}
	\renewcommand{\labelsep}{none}
	\caption{\textbf{Calculated loss rate constants $L_3(\textrm{v},R)$ versus calculated binding energy $E_b(\textrm{v}, R)$ for our model with $N_s=6$ $s$-wave bound states.}
		The dashed lines are to guide the eye between the different rotational levels of vibrational states v $=$ $-2$, $-3$, $-4$, $-5$ and $-6$; the rotational quantum numbers increase in steps of 2 from the lowest energy $R=0$ level for each v.  The solid red line shows a line varying proportional to $1/\sqrt{E_b}$.  }
	\label{Supfig:Fig7}
\end{figure*}

Our calculations verify that $L_3(\textrm{v},R)$ varies by less than 2\% as a function of collision energy $E_{col}$ in the range of 0 to $ 1\:\mu\text{K}\times k_B$.
 Our total $L_3$ value is converged with respect to  $N_s$. We obtain $L_3=1.5\times10^{-29}\:\text{cm}^6\,\text{s}^{-1}$ for $N_s=1$ and $L_3=1.0\times10^{-29}\:\text{cm}^6\,\text{s}^{-1}$ for $N_s=2,3,4,5,$ and $6$. Our calculated value of $L_3=1.0\times10^{-29}\:\text{cm}^6\,\text{s}^{-1}$ is a factor of 4 less than the one measured by Burt \textit{et al.} \cite{Burt1997} and a factor of 2 less than their lower error range. Clearly more work in experiment and theory is required to establish an accurate magnitude of $L_3$ with less uncertainty. Therefore, in order to eliminate the differences in magnitude and compare the relative population distributions between theory and experiment, the calculated values have been scaled by a factor of 4.3 in Fig.$\:$ 5 to normalize them to the magnitude reported by \cite{Burt1997}.

We have also tested the possible effect of pairwise non-additive long-range Axilrod-Teller corrections \cite{Axilrod1943} to the 3-body potential. Assuming a plausible form and magnitude for such corrections shows that they change the total recombination rate less than a few percent. We have observed that a larger effect on product distributions occurs for molecular states with larger binding energies. While these effects can be larger than the few percent level for the most-deeply bound molecular states considered in the calculations, we have seen no significant qualitative change in the product distributions of the most-weakly bound molecular states.

\vspace{-0.4cm}
\subsection*{13. Error estimation for population distribution}
\vspace{-0.4cm}
In Fig.$\:$5 the error bars represent the statistical uncertainties of the ion signals. However, there are also other sources of error. Due to an imperfect atom number calibration in our setup of up to $30\%$ there is an uncertainty of $60\%$ in the global normalization of the flux into the molecular quantum states. Furthermore, for the determination of $\eta_1$ we assume the relaxation rate constant of the product molecule to be $K_{\text{rel}} = 10^{-10}\:\text{cm}^{3}\,\text{s}^{-1}$ (see \cite{SMukaiyama2004,SStaanum2006,SZahzam2006,SQuemener2007,SQuemener2005}), a value which could easily be off by a factor of two.  While for the levels with $\text{v} = -2,\dots,-5$  this would entail only a small correction of up to a few percent, for the level v = -1 this correction could reach  35\%.
\end{section}

\newpage
\bibliographystyle{plain}

\begin{thebibliography}{99}
	\expandafter\ifx\csname url\endcsname\relax
	\def\url#1{\texttt{#1}}\fi
	\expandafter\ifx\csname urlprefix\endcsname\relax\def\urlprefix{URL }\fi
	\providecommand{\bibinfo}[2]{#2}
	\providecommand{\eprint}[2][]{\url{#2}}
%
	\makeatletter
    \renewcommand{\@biblabel}[1]{#1.} 
	\bibitem{Suno2009}
	\bibinfo{author}{H. Suno},
	\bibinfo{author}{B.~D. Esry},
	\newblock \bibinfo{title}{Three-body recombination in cold helium-helium-alkali-metal-atom collisions}.
	\newblock \emph{\bibinfo{journal}{Phys. Rev. A}}
	\textbf{\bibinfo{volume}{80}}, \bibinfo{pages}{062702}
	(\bibinfo{year}{2009}).
	
	\bibitem{Wang2011}
	\bibinfo{author}{Y. Wang}, \bibinfo{author}{J.~P. D'Incao},
	\bibinfo{author}{B.~D. Esry},
	\newblock \bibinfo{title}{Cold three-body collisions in hydrogen-hydrogen-alkali-metal atomic systems}.
	\newblock \emph{\bibinfo{journal}{Phys. Rev. A}}
	\textbf{\bibinfo{volume}{83}}, \bibinfo{pages}{032703}
	(\bibinfo{year}{2011}).

\bibitem{Croft2017}
	\bibinfo{author}{J.~F.~E. Croft, \textit{et al.}},
	\newblock \bibinfo{title}{Quantum reactive scattering of ultracold atoms and molecules: universality and chaotic dynamics}.
	\newblock  \bibinfo{pages}{Preprint at https://arxiv.org/abs/1701.09090}
	(\bibinfo{year}{2017}).

	\bibitem{Wang2011b}
	\bibinfo{author}{J. Wang}, \bibinfo{author}{J.~P. D'Incao },
	\bibinfo{author}{C.~H. Greene},
	\newblock \bibinfo{title}{Numerical study of three-body recombination for systems with many bound states}.
	\newblock \emph{\bibinfo{journal}{Phys. Rev. A}}
	\textbf{\bibinfo{volume}{84}}, \bibinfo{pages}{052721}
	(\bibinfo{year}{2011}).

	\bibitem{Gerrity1984}
	\bibinfo{author}{D.~P. Gerrity, J.~J. Valentini},
	\newblock \bibinfo{title}{Experimental study of the dynamics of the $\text{H}+\text{D}_2\rightarrow \text{HD}+\text{D}$ reaction at collision energies of 0.55 and 1.30$\:$eV}.
	\newblock \emph{\bibinfo{journal}{J. Chem. Phys.}}
	\textbf{\bibinfo{volume}{81}}, \bibinfo{pages}{1298-1313}
	(\bibinfo{year}{1984}).

    \bibitem{Neuhauser1992}
	\bibinfo{author}{D. Neuhauser, \textit{et al.}}
	\newblock \bibinfo{title}{State-to-State rates for the $\text{D} + \text{H}_2 (\text{v} = 1, j = 1) \rightarrow \text{HD}(\text{v}',j' ) + \text{H}$ reaction: predictions and measurements}.
	\newblock \emph{\bibinfo{journal}{Science}}
	\textbf{\bibinfo{volume}{257}}, \bibinfo{pages}{519-522}
	(\bibinfo{year}{1992}).

    \bibitem{Kitsopoulos1993}
	\bibinfo{author}{T.~N. Kitsopoulos, M.~A. Buntine, D.~P. Baldwin, R.~N. Zare, D.~W. Chandler,}
	\newblock \bibinfo{title}{Reaction product imaging: the $\text{H} + \text{D}_2$ reaction}.
	\newblock \emph{\bibinfo{journal}{Science}}
	\textbf{\bibinfo{volume}{260}}, \bibinfo{pages}{1605-1610}
	(\bibinfo{year}{1993}).

    \bibitem{Becker1995}
	\bibinfo{author}{S. Becker, C. Braatz, J. Lindner, E. Tiemann,}
	\newblock \bibinfo{title}{Investigation of the predissociation of SO$_2$: state selective detection of the SO and O fragments}.
	\newblock \emph{\bibinfo{journal}{Chem. Phys.}}
	\textbf{\bibinfo{volume}{196}}, \bibinfo{pages}{275-291}
	(\bibinfo{year}{1995}).

    \bibitem{Weber2003}
	\bibinfo{author}{T. Weber}, \bibinfo{author}{J. Herbig,} \bibinfo{author}{M. Mark}, \bibinfo{author}{H.-C. N\"{a}gerl},
	\bibinfo{author}{R. Grimm},
	\newblock \bibinfo{title}{Three-body recombination at large scattering lengths in an ultracold atomic gas}.
	\newblock \emph{\bibinfo{journal}{Phys. Rev. Lett.}}
	\textbf{\bibinfo{volume}{91}}, \bibinfo{pages}{123201}
	(\bibinfo{year}{2003}).
		
	\bibitem{Jochim2003}
	\bibinfo{author}{S. Jochim, \textit{et al.}}
	\newblock \bibinfo{title}{Pure gas of optically trapped molecules created from fermionic atoms}.
	\newblock \emph{\bibinfo{journal}{Phys. Rev. Lett.}}
	\textbf{\bibinfo{volume}{91}}, \bibinfo{pages}{240402}
	(\bibinfo{year}{2003}).

    \bibitem{Rui2017}
	\bibinfo{author}{J. Rui, \textit{et al.}}
	\newblock \bibinfo{title}{Controlled state-to-state atom-exchange reaction in an ultracold atom-dimer mixture}.
	\newblock \emph{\bibinfo{journal}{Nat. Phys.}} DOI 10.1038/nphys4095 (\bibinfo{year}{2017}).

    \bibitem{Lyon2017}
	\bibinfo{author}{M. Lyon, S.~L. Rolston,}
	\newblock \bibinfo{title}{Ultracold neutral plasmas}.
	\newblock \emph{\bibinfo{journal}{Rep. Prog. Phys.}}
	\textbf{\bibinfo{volume}{80}}, \bibinfo{pages}{017001}
	(\bibinfo{year}{2017}).

    \bibitem{Baulch1992}
	\bibinfo{author}{D. L. Baulch, \textit{et al.}}
	\newblock \bibinfo{title}{Evaluated kinetic data for combustion modelling}.
	\newblock \emph{\bibinfo{journal}{J. Chem. Phys. Ref. Data}}
	\textbf{\bibinfo{volume}{21}}, \bibinfo{pages}{411-734}
	(\bibinfo{year}{1992}).

    \bibitem{Brown1999}
	\bibinfo{author}{S. S. Brown, R. K. Talukdar, A. R. Ravishankara,}
	\newblock \bibinfo{title}{Rate constants for the reaction OH+NO$_2$+M $\rightarrow$ HNO$_3$+M under atmospheric conditions}.
	\newblock \emph{\bibinfo{journal}{Chem. Phys. Lett.}}
	\textbf{\bibinfo{volume}{299}}, \bibinfo{pages}{277-284}
	(\bibinfo{year}{1999}).

    \bibitem{Turk2011}
	\bibinfo{author}{M. J. Turk, \textit{et al.}}
	\newblock \bibinfo{title}{Effects of varying the three-body molecular hydrogen formation rate in primordial star formation}.
	\newblock \emph{\bibinfo{journal}{Astrophys. J.}}
	\textbf{\bibinfo{volume}{726}}, \bibinfo{pages}{55}
	(\bibinfo{year}{2011}).

    \bibitem{Forrey2013}
	\bibinfo{author}{R. C. Forrey,}
	\newblock \bibinfo{title}{Rate of formation of hydrogen molecules by three-body recombination during primordial star formation}.
	\newblock \emph{\bibinfo{journal}{Astrophys. Lett.}}
	\textbf{\bibinfo{volume}{773}}, \bibinfo{pages}{2}
	(\bibinfo{year}{2013}).

    \bibitem{Moerdijk1996}
	\bibinfo{author}{A.~J. Moerdijk}, \bibinfo{author}{H.~M.~J.~M. Boesten,}
	\bibinfo{author}{B.~J. Verhaar,}
	\newblock \bibinfo{title}{Decay of trapped ultracold alkali atoms by recombination}.
	\newblock \emph{\bibinfo{journal}{Phys. Rev. A}}
	\textbf{\bibinfo{volume}{53}}, \bibinfo{pages}{916-920}
	(\bibinfo{year}{1996}).

   \bibitem{Fedichev1996}
	\bibinfo{author}{P.~O. Fedichev}, \bibinfo{author}{M.~W. Reynolds,}
	\bibinfo{author}{G.~V. Shlyapnikov,}
	\newblock \bibinfo{title}{Three-body recombination of ultracold atoms to a weakly bound $s$ level}.
	\newblock \emph{\bibinfo{journal}{Phys. Rev. Lett.}}
	\textbf{\bibinfo{volume}{77}}, \bibinfo{pages}{2921-2924}
	(\bibinfo{year}{1996}).

	\bibitem{Soeding1999}
	\bibinfo{author}{J. S\"{o}ding, \textit{et al.}}
	\newblock \bibinfo{title}{Three-body decay of a rubidium Bose-Einstein condensate}.
	\newblock \emph{\bibinfo{journal}{Appl. Phys. B}}
	\textbf{\bibinfo{volume}{69}}, \bibinfo{pages}{257-261}
	(\bibinfo{year}{1999}).	

    \bibitem{Quemener2012}
	\bibinfo{author}{G. Qu\'{e}m\'{e}ner}, \bibinfo{author}{P.~S. Julienne,}
	\newblock \bibinfo{title}{Ultracold molecules under control!}
	\newblock \emph{\bibinfo{journal}{Chem.  Rev.}}
	\textbf{\bibinfo{volume}{112}}, \bibinfo{pages}{4949-5011}
	(\bibinfo{year}{2012}).

    \bibitem{Chin2010}
	\bibinfo{author}{C. Chin}, \bibinfo{author}{R. Grimm}, \bibinfo{author}{P. Julienne,}
	\bibinfo{author}{E. Tiesinga,}
	\newblock \bibinfo{title}{Feshbach resonances in ultracold gases}.
	\newblock \emph{\bibinfo{journal}{Rev. Mod. Phys.}}
	\textbf{\bibinfo{volume}{82}}, \bibinfo{pages}{1225-1286}
	(\bibinfo{year}{2010}).
	
	\bibitem{Burt1997}
	\bibinfo{author}{E.~A. Burt, \textit{et al.}}
	\newblock \bibinfo{title}{Coherence, correlations, and collisions: what one learns about Bose-Einstein condensates from their decay}.
	\newblock \emph{\bibinfo{journal}{Phys. Rev. Lett.}}
	\textbf{\bibinfo{volume}{79}}, \bibinfo{pages}{337-340}
	(\bibinfo{year}{1997}).
	
	\bibitem{Haerter2012}
	\bibinfo{author}{A. H\"{a}rter, \textit{et al.}}
	\newblock \bibinfo{title}{Single ion as a three-body reaction center in an ultracold atomic gas}.
	\newblock \emph{\bibinfo{journal}{Phys. Rev. Lett.}} \textbf{\bibinfo{volume}{109}},
	\bibinfo{pages}{123201} (\bibinfo{year}{2012}).
	
	\bibitem{Kruekow2016}
	\bibinfo{author}{A. Kr\"{u}kow, \textit{et al.}}
	\newblock \bibinfo{title}{Energy scaling of cold atom-atom-ion three-body recombination}.
	\newblock \emph{\bibinfo{journal}{Phys. Rev. Lett.}}
	\textbf{\bibinfo{volume}{116}}, \bibinfo{pages}{193201}
	(\bibinfo{year}{2016}).
	
	\bibitem{Shotan2014}
	\bibinfo{author}{Z. Shotan}, \bibinfo{author}{O. Machtey}, \bibinfo{author}{S. Kokkelmans,}
	\bibinfo{author}{L. Khaykovich,}
	\newblock \bibinfo{title}{Three-body recombination at vanishing scattering lengths in an ultracold Bose gas}.
	\newblock \emph{\bibinfo{journal}{Phys. Rev. Lett.}} \textbf{\bibinfo{volume}{113}},
	\bibinfo{pages}{053202} (\bibinfo{year}{2014}).
	
	\bibitem{Kraemer2006}
	\bibinfo{author}{T. Kraemer, \textit{et al.}.}
	\newblock \bibinfo{title}{Evidence for Efimov quantum states in an ultracold
			gas of caesium atoms}.
	\newblock \emph{\bibinfo{journal}{Nature}}
	\textbf{\bibinfo{volume}{440}}, \bibinfo{pages}{315-318}
	(\bibinfo{year}{2006}).

\bibitem{Bedaque2000}
	\bibinfo{author}{P.~F. Bedaque}, \bibinfo{author}{E. Braaten,}
	\bibinfo{author}{H.-W. Hammer,}
	\newblock \bibinfo{title}{Three-body recombination in Bose gases with large scattering length}.
	\newblock \emph{\bibinfo{journal}{Phys. Rev. Lett.}}
	\textbf{\bibinfo{volume}{85}}, \bibinfo{pages}{908-911}
	(\bibinfo{year}{2000}).
	
	\bibitem{Petrov2003}
	\bibinfo{author}{D.~S. Petrov,}
	\newblock \bibinfo{title}{Three-body problem in Fermi gases with short-range interparticle interaction}.
	\newblock \emph{\bibinfo{journal}{Phys. Rev. A}}
	\textbf{\bibinfo{volume}{67}}, \bibinfo{pages}{010703(R)}
	(\bibinfo{year}{2003}).
	
	\bibitem{Esry1999}
	\bibinfo{author}{B.~D. Esry}, \bibinfo{author}{C.~H. Greene,}
	\bibinfo{author}{J.~P. Burke,}
	\newblock \bibinfo{title}{Recombination of three atoms in the ultracold limit}.
	\newblock \emph{\bibinfo{journal}{Phys. Rev. Lett.}} \textbf{\bibinfo{volume}{83}},
	\bibinfo{pages}{1751-1754} (\bibinfo{year}{1999}).
	
	\bibitem{Incao2005}
	\bibinfo{author}{J.~P. D'Incao,}
	\bibinfo{author}{B.~D. Esry,}
	\newblock \bibinfo{title}{Scattering length scaling laws for ultracold three-body collisions}.
	\newblock \emph{\bibinfo{journal}{Phys. Rev. Lett.}}
	\textbf{\bibinfo{volume}{94}}, \bibinfo{pages}{213201}
	(\bibinfo{year}{2005}).
	
	\bibitem{Petrov2004}
	\bibinfo{author}{D.~S. Petrov,}
	\newblock \bibinfo{title}{Three-boson problem near a narrow Feshbach resonance}.
	\newblock \emph{\bibinfo{journal}{Phys. Rev. Lett.}}
	\textbf{\bibinfo{volume}{93}}, \bibinfo{pages}{143201}
	(\bibinfo{year}{2004}).
	
	\bibitem{Perez2014}
	\bibinfo{author}{J. P\'{e}r\'{e}z-R\'{i}os}, \bibinfo{author}{S. Ragole}, \bibinfo{author}{J. Wang,}
	\bibinfo{author}{C.~H. Greene,}
	\newblock \bibinfo{title}{Comparison of classical and quantal calculations of helium three-body recombination}.
	\newblock \emph{\bibinfo{journal}{J. Chem. Phys.}}
	\textbf{\bibinfo{volume}{140}}, \bibinfo{pages}{044307}
	(\bibinfo{year}{2014}).
	
	\bibitem{Wang2014}
	\bibinfo{author}{Y. Wang,}
	\bibinfo{author}{P.~S. Julienne,}
	\newblock \bibinfo{title}{Universal van der Waals physics for three cold
			atoms near Feshbach resonances}.
	\newblock \emph{\bibinfo{journal}{Nat. Phys.}}
	\textbf{\bibinfo{volume}{10}}, \bibinfo{pages}{768-773}
	(\bibinfo{year}{2014}).	
	
	\bibitem{Braaten2006}
	\bibinfo{author}{E. Braaten,}
	\bibinfo{author}{H.-W. Hammer,}
	\newblock \bibinfo{title}{Universality in few-body systems with large scattering length}.
	\newblock \emph{\bibinfo{journal}{Phys. Rep.}}
	\textbf{\bibinfo{volume}{428}}, \bibinfo{pages}{259-390}
	(\bibinfo{year}{2006}).

    \bibitem{Nesbitt2012}
	\bibinfo{author}{D. Nesbitt,} \newblock \bibinfo{title}{Toward state-to-state dynamics in ultracold collisions: lessons from high-resolution spectroscopy of weakly bound molecular complexes}.
	\newblock \emph{\bibinfo{journal}{Chem. Rev.}}
	\textbf{\bibinfo{volume}{112}}, \bibinfo{pages}{5062-5072}
	(\bibinfo{year}{2012}).

    \bibitem{Gonz2014}
	\bibinfo{author}{M.~L. Gonz\'{a}lez-Mart\'{i}nez, O. Dulieu, P. Larr\'{e}garay, L. Bonnet,}
	\newblock \bibinfo{title}{Statistical product distributions for ultracold reactions in external fields}.
	\newblock \emph{\bibinfo{journal}{Phys. Rev. A}}
	\textbf{\bibinfo{volume}{90}}, \bibinfo{pages}{052716}
	(\bibinfo{year}{2014}).
	
	    \bibitem{Supp}
	\bibinfo{author}{Materials and methods are available as supplementary materials.}
	
		\bibitem{Drozdova2013}
	\bibinfo{author}{A. Drozdova, \textit{et al.}}
	\newblock \bibinfo{title}{Fourier transform spectroscopy and extended deperturbation treatment of the spin-orbit-coupled $A^1\Sigma_u^+$ and $b^3\Pi_u$ states of the Rb$_2$ molecule}.
	\newblock \emph{\bibinfo{journal}{Phys. Rev. A}}
	\textbf{\bibinfo{volume}{88,}} \bibinfo{pages}{022504} (\bibinfo{year}{2013}).
	
	\bibitem{Deiss2015}
	\bibinfo{author}{M. Dei{\ss}}, \bibinfo{author}{B. Drews}, \bibinfo{author}{J. Hecker Denschlag,}
	\bibinfo{author}{E. Tiemann,}
	\newblock \bibinfo{title}{Mixing of $0^+$ and $0^-$ observed in the hyperfine and Zeeman structure of ultracold Rb$_{2}$ molecules}.
	\newblock \emph{\bibinfo{journal}{New J. Phys.}}
	\textbf{\bibinfo{volume}{17,}} \bibinfo{pages}{083032}
	(\bibinfo{year}{2015}).
		
   \bibitem{Jones2006}
	\bibinfo{author}{K.~M. Jones}, \bibinfo{author}{E. Tiesinga}, \bibinfo{author}{P.~D. Lett,}
	\bibinfo{author}{P.~S. Julienne,}
	\newblock \bibinfo{title}{Ultracold photoassociation spectroscopy: Long-range molecules and atomic scattering}.
	\newblock \emph{\bibinfo{journal}{Rev. Mod. Phys.}}
	\textbf{\bibinfo{volume}{78}}, \bibinfo{pages}{483-535}
	(\bibinfo{year}{2006}).

    \bibitem{Haerter2013}
	\bibinfo{author}{A. H\"{a}rter, \textit{et al.}}
	\newblock \bibinfo{title}{Population distribution of product states following three-body recombination in an ultracold atomic gas}.
	\newblock \emph{\bibinfo{journal}{Nat. Phys.}} \textbf{\bibinfo{volume}{9}},
	\bibinfo{pages}{512--517} (\bibinfo{year}{2013}).
	
	
	
		\bibitem{SStrauss2010}
	\bibinfo{author}{C. Strauss, \textit{et al.}},
	\newblock \bibinfo{title}{Hyperfine, rotational and vibrational structure of of the a $^3\Sigma^+_u$ state of $^{87}$Rb$_{2}$ molecules}.
	\newblock \emph{\bibinfo{journal}{Phys. Rev. A}}
	\textbf{\bibinfo{volume}{82}}, \bibinfo{pages}{052514}
	(\bibinfo{year}{2010}).

	\bibitem{SDrews2017}
	\bibinfo{author}{B. Drews}, \bibinfo{author}{M. Dei{\ss}}, \bibinfo{author}{J. Wolf}, \bibinfo{author}{E. Tiemann,}  \bibinfo{author}{J. Hecker Denschlag},
	\newblock \bibinfo{title}{Level structure of deeply bound levels of the c$^3\Sigma^+_g$ state of $^{87}$ Rb$_{2}$ molecules}.
\newblock  \bibinfo{pages}{Preprint at https://arxiv.org/abs/1703.07752}
	(\bibinfo{year}{2017}).
		
		\bibitem{SGuan2013}
	\bibinfo{author}{Y. Guan}, \bibinfo{author}{et al.},
	\newblock \bibinfo{title}{Updated potential energy function of the Rb$_{2}$ a$^3\Sigma^+_u$ state in the attractive and repulsive regions determined from its joint analysis with the 2$^3\Pi_{0g}$ state}.
	\newblock \emph{\bibinfo{journal}{J. Chem. Phys.}}
	\textbf{\bibinfo{volume}{139}}, \bibinfo{pages}{144303}
	(\bibinfo{year}{2013}).
	
	\bibitem{Haerter2013b}
	\bibinfo{author}{A. H\"{a}rter}, \bibinfo{author}{A. Kr\"{u}kow}, \bibinfo{author}{A. Brunner,}
	\bibinfo{author}{J. Hecker Denschlag,}
	\newblock \bibinfo{title}{Minimization of ion micromotion using ultracold atomic probes}.
	\newblock \emph{\bibinfo{journal}{Appl. Phys. Lett.}}
	\textbf{\bibinfo{volume}{102}}, \bibinfo{pages}{221115}
	(\bibinfo{year}{2013}).
	
   \bibitem{SMukaiyama2004}
	\bibinfo{author}{T. Mukaiyama}, \bibinfo{author}{J.~R. Abo-Shaeer},
    \bibinfo{author}{K. Xu}, \bibinfo{author}{J.~K. Chin,}
    \bibinfo{author}{W. Ketterle,}
	\newblock \bibinfo{title}{Dissociation and decay of ultracold sodium molecules}.
	\newblock \emph{\bibinfo{journal}{Phys. Rev. Lett.}}
	\textbf{\bibinfo{volume}{92}}, \bibinfo{pages}{180402}
	(\bibinfo{year}{2004}).

    \bibitem{SStaanum2006}
	\bibinfo{author}{P. Staanum}, \bibinfo{author}{S.~D. Kraft},
    \bibinfo{author}{J. Lange}, \bibinfo{author}{R. Wester,}
    \bibinfo{author}{M. Weidem\"{u}ller,}
	\newblock \bibinfo{title}{Experimental
investigation of ultracold atom-molecule collisions}.
	\newblock \emph{\bibinfo{journal}{Phys. Rev. Lett.}}
	\textbf{\bibinfo{volume}{96}}, \bibinfo{pages}{023201}
	(\bibinfo{year}{2006}).

    \bibitem{SZahzam2006}
	\bibinfo{author}{N. Zahzam}, \bibinfo{author}{T. Vogt},
    \bibinfo{author}{M. Mudrich}, \bibinfo{author}{D. Comparat},
    \bibinfo{author}{P. Pillet,}
	\newblock \bibinfo{title}{Atom-molecule
collisions in an optically trapped gas}.
	\newblock \emph{\bibinfo{journal}{Phys. Rev. Lett.}}
	\textbf{\bibinfo{volume}{96}}, \bibinfo{pages}{023202}
	(\bibinfo{year}{2006}).

    \bibitem{SQuemener2007}
	\bibinfo{author}{G. Qu\'{e}m\'{e}ner}, \bibinfo{author}{J.-M. Launay,} \bibinfo{author}{P. Honvault,}
	\newblock \bibinfo{title}{Ultracold collisions between Li atoms and Li$_2$
diatoms in high vibrational states.}.
	\newblock \emph{\bibinfo{journal}{Phys. Rev. A}}
	\textbf{\bibinfo{volume}{75}}, \bibinfo{pages}{050701(R)}
	(\bibinfo{year}{2007}).

    \bibitem{SQuemener2005}
	\bibinfo{author}{G. Qu\'{e}m\'{e}ner, \textit{et al.}},
	\newblock \bibinfo{title}{Ultracold quantum dynamics: Spin-polarised $\text{K}+\text{K}_2$ collisions
with three identical bosons or fermions}.
	\newblock \emph{\bibinfo{journal}{Phys. Rev. A}}
	\textbf{\bibinfo{volume}{71}}, \bibinfo{pages}{032722}
	(\bibinfo{year}{2005}).	
		
   \bibitem{Mehta2009}
   \bibinfo{author}{N.~P. Mehta}, \bibinfo{author}{S.~T. Rittenhouse}, \bibinfo{author}{J.~P. D'Incao}, \bibinfo{author}{J. von Stecher,}
   \bibinfo{author}{C.~H. Greene,}
    \newblock \bibinfo{title}{General theoretical description of $N$-body recombination}.
   \newblock \emph{\bibinfo{journal}{Phys. Rev. Lett.}}
   \textbf{\bibinfo{volume} {103}},
   \bibinfo{pages}{153201}
   (\bibinfo{year}{2009}).

 \bibitem{Wang2012}
	\bibinfo{author}{J. Wang}, \bibinfo{author}{J.~P. D'Incao}, \bibinfo{author}{Y. Wang,}
	\bibinfo{author}{C.~H. Greene,}
	\newblock \bibinfo{title}{Universal three-body recombination via resonant $d$-wave interactions}.
	\newblock \emph{\bibinfo{journal}{Phys. Rev. A}}
	\textbf{\bibinfo{volume}{86}}, \bibinfo{pages}{062511}
	(\bibinfo{year}{2012}).
	
	
	\bibitem{Axilrod1943}
	\bibinfo{author}{B.~M. Axilrod,} \bibinfo{author}{E. Teller,}
	\newblock \bibinfo{title}{Interaction of the van der Waals type between three atoms}.
	\newblock \emph{\bibinfo{journal}{J. Chem. Phys.}}
	\textbf{\bibinfo{volume}{11}}, \bibinfo{pages}{299-300}
	(\bibinfo{year}{1943}).
	
	    \bibitem{Ack}
	\bibinfo{author}{\textbf{Acknowledgments:} This work was supported by the German research foundation Deutsche Forschungsgemeinschaft (DFG) within SFB/TRR21 and Grant DE 510/2-1.
A.K. acknowledges support from the Carl Zeiss foundation.
E.T. acknowledges support from the Minister of Science and Culture of Lower Saxony, Germany, by providing a Niedersachsenprofessur. J.P.D. acknowledges NSF funding (Grant PHY-1607204). J.H.D., P.S.J., E.T., B.P.R., Y.W., and J.P.D. acknowledge support from several KITP programs (NSF Grant PHY-1125915). J.H.D. would like to thank David Nesbit for insightful encouragement.\\
\textbf{Author Contributions} J.W., A.K., M.D. have carried out the experiments. E.T. has carried out coupled-channel calculations for Rb$_2$. B.P.R., J.P.D., Y.W., and P.S.J. have set up, implemented, and analyzed the three-body calculations. J.H.D. supervised the project. All authors have contributed to the analysis of the experiment and to the writing of the manuscript.\\
}	
\end{thebibliography}

\end{document}